\documentclass[aps,pra,twocolumn,superscriptaddress, longbibliography]{revtex4-2}

\usepackage{graphicx,xcolor}
\usepackage{amsmath,amsfonts,amssymb}

\usepackage[]{siunitx}
\usepackage{chemformula}

\usepackage{hyperref}

\begin{document}

\title{Comparing Transmission- and Epi-BCARS:\linebreak A Transnational Round Robin on Solid State Materials}



\author{Franz Hempel}
\email[email]{: franz.hempel{@}tu-dresden.de}
\affiliation{Institut für Angewandte Physik, Technische Universität Dresden, 01062 Dresden, Germany}

\author{Federico Vernuccio}
\email[email]{: federico.vernuccio@polimi.it}
\affiliation{Physics Department, Politecnico di Milano, 20133 Milano, Italy}

\author{Lukas König}
\affiliation{Institut für Angewandte Physik, Technische Universität Dresden, 01062 Dresden, Germany}

\author{Robin Buschbeck}
\affiliation{Institut für Angewandte Physik, Technische Universität Dresden, 01062 Dresden, Germany}

\author{Michael Rüsing}
\affiliation{Institut für Angewandte Physik, Technische Universität Dresden, 01062 Dresden, Germany}

\author{Giulio Cerullo}
\affiliation{Physics Department, Politecnico di Milano, 20133 Milano, Italy}

\author{Dario Polli}
\affiliation{Physics Department, Politecnico di Milano, 20133 Milano, Italy}

\author{Lukas M. Eng}
\affiliation{Institut für Angewandte Physik, Technische Universität Dresden, 01062 Dresden, Germany}
\affiliation{ct.qmat: Dresden-Würzburg Cluster of Excellence—EXC 2147, Technische Universität Dresden, 01062 Dresden, Germany}

\date{\today}

\begin{abstract}

Broadband coherent anti-Stokes Raman scattering (BCARS) is an advanced Raman spectroscopy method that combines the spectral sensitivity of spontaneous Raman scattering (SR) with the increased signal intensity of single-frequency coherent Raman techniques.
These two features make BCARS particularly suitable for ultra-fast imaging of heterogeneous samples, as already shown in biomedicine.
Recent studies demonstrated that BCARS also shows exceptional spectroscopic capabilities when inspecting crystalline materials like lithium niobate and lithium tantalate, and can be used for fast imaging of ferroelectric domain walls.
These results strongly suggest the extension of BCARS towards new imaging applications like mapping defects, strain, or dopant levels, similar to standard SR imaging.
Despite these advantages, BCARS suffers from a spurious and chemically unspecific non-resonant background (NRB) that distorts and shifts the Raman peaks.
Post-processing numerical algorithms are then used to remove the NRB and to obtain spectra comparable to SR results.
Here, we show the reproducibility of BCARS by conducting an internal Round Robin with two different BCARS experimental setups, comparing the results on different crystalline materials of increasing structural complexity: diamond, 6H-SiC,  KDP, and KTP.
First, we compare the detected and phase-retrieved signals, the setup-specific NRB-removal steps, and the mode assignment.
Subsequently, we demonstrate the versatility of BCARS by showcasing how the selection of pump wavelength, pulse width, and detection geometry can be tailored to suit the specific objectives of the experiment.
Finally, we compare and optimize measurement parameters for the high-speed, hyperspectral imaging of ferroelectric domain walls in lithium niobate.

\end{abstract}

\keywords{broadband coherent anti-Stokes Raman scattering, BCARS, CARS, solid state, single crystals, phase-retrieval, phase matching}
\maketitle


\section{Introduction}

Spontaneous Raman scattering (SR) is a powerful vibrational spectroscopy technique that provides rich chemical information on materials and biomedical samples in a non-destructive and label-free way \cite{Vanna2022, turrell1996}.
Due to its high sensitivity to changes in the crystal structure, it is particularly useful to detect changes like defects \cite{Fontana2015}, strain \cite{hayazawa2007}, and doping \cite{prawer2004}.
However, one drawback is the low scattering efficiency of the Raman effect, which leads to long pixel dwell times and prevents image acquisition with high spatial resolution.
Coherent Raman scattering (CRS) dramatically increases the acquisition speed thanks to the nonlinear nature of the optical interaction between the sample and the impinging electric fields, which generate a vibrational coherence \cite{Polli2018}.
In its simplest configuration, narrowband CRS typically employs two picosecond pulses, the pump and the Stokes, at frequency $\omega_p$ and $\omega_s$, respectively, interacting with the sample. The frequency difference is tuned to match and interrogate a chosen vibrational mode $\Omega = \omega_p-\omega_s$.
Among the CRS techniques, Coherent Anti-Stokes Raman Scattering (CARS) has emerged as an advanced, yet quite simple-to-implement method for high-speed spectroscopy and imaging \cite{cheng2016}.
The CARS signal derives from a four-wave mixing process at the sample: first, interactions with the pump and Stokes fields lead to the population of vibrational states. Then, a probe field reads these states, generating the blue-shifted anti-Stokes component at frequency $\omega_{as}$. In most implementations, the pump and probe field are frequency degenerate since they are delivered by the same train of pulses, leading to $\omega_{as} = 2 \omega_p - \omega_s$.
However, narrowband CARS does not provide the same chemical specificity as SR since it probes only a single frequency at a time.
The solution is Broadband CARS (BCARS), which combines a narrowband pump with a broadband Stokes pulse and probes a wide range of vibrational frequencies of the sample. BCARS combines the increase in signal intensity of CRS techniques with the rich spectroscopic information provided by SR \cite{Polli2018}.

One prominent feature of CARS is the influence of the so-called non-resonant background (NRB) signal.
The NRB results from non-resonant four-wave mixing processes in the sample and its surroundings.
It coherently mixes with the resonant response of the investigated material, which results in enhanced signal intensity, but also in the distortion of both peak shape and peak position.
Several solutions may be adopted to suppress this chemically unspecific signal, like time-resolved detection \cite{Kolesnichenko2019}, polarization CARS \cite{Cheng:01}, Fourier-transform CARS \cite{Cui:06}, or angle-resolved measurements of anisotropic vibrational modes \cite{cheng2002}. All these methods, however, increase the setup complexity and lead to signal degradation. For BCARS, the NRB can be removed using analytical approaches like the maximum entropy method \cite{Vartiainen1992}, the time-domain Kramers-Kronig (KK) transformation  \cite{Camp2016}, or deep-learning-based methods \cite{Valensise2020, Houhou2020, Wang2022}.

Today, BCARS is broadly applied in the biomedical field \cite{Delacadena2022, Cicerone2016} for cancer research\cite{Vernuccio2023}, label-free imaging, and for sample sizes down to nanoparticles \cite{Pope2014}. In addition, recent works have demonstrated excellent performances of BCARS for the investigation of ferroelectric materials, such as \ch{LiNbO_3} and \ch{LiTaO_3}\cite{Hempel2021} as well as imaging of single ferroelectric domain walls \cite{Reitzig2022}.

In biomedical imaging, the strong signal allows for fast and label-free imaging with sub-ms acquisition times. The intensity of specific bands or modes is used to contrast different types of tissue \cite{evans2008, petrov2021, Zhang2015}.
The focus is different for crystalline materials.
Knowing and retrieving the exact peak parameters of the vibrational modes is crucial in solid-state physics, as changes in the material's structure, like strain or defects, affect the Raman spectrum only slightly.
Small shifts in peak position or width broadening indicate defects, strain, dopants, number of layers, and structural changes \cite{Fontana2015, hayazawa2007, prawer2004, late2011, ferrari2006}.
It is critical that differences in experimental setups used do not change the results of these precise measurements.

Interlaboratory setup comparisons, i.e., Round Robin investigations, have been made for SR spectroscopy \cite{Guo2020}, Auger electron spectroscopy \cite{Powell1982}, mass spectroscopy \cite{Bristow2003, Leymarie2013} and tip-enhanced Raman spectroscopy \cite{Blum2013}, showing the reproducibility of their respective techniques and laying the basis for comparative research and widespread application.

To the best of our knowledge, only one comparison between BCARS spectra of glycerol taken with two different setups has been reported in literature \cite{Camp2016}. Reproducibility is an essential step towards the broader use of this technology. One key aspect of that is the strong setup dependence of the NRB due to differences in the spectral profiles of the used laser sources and detector efficiencies. It is of great importance that these differences are taken into account in the NRB removal process.

Here, we compare two BCARS setups with greatly different parameters, evaluating their reproducibility when measuring and employing NRB suppression algorithms on a range of materials.
In addition, we compare measurement parameters like the forward and backward (epi) detection directions and focal depth dependence for the application to four crystalline materials of increasing structural complexity. Finally, we optimize the high-speed imaging of ferroelectric domain walls.

\section{BCARS Spectroscopy}

\subsection{Inelastic Light-Matter Interaction}

\begin{figure}[htbp]
	\centering
	\includegraphics[width=0.4\textwidth]{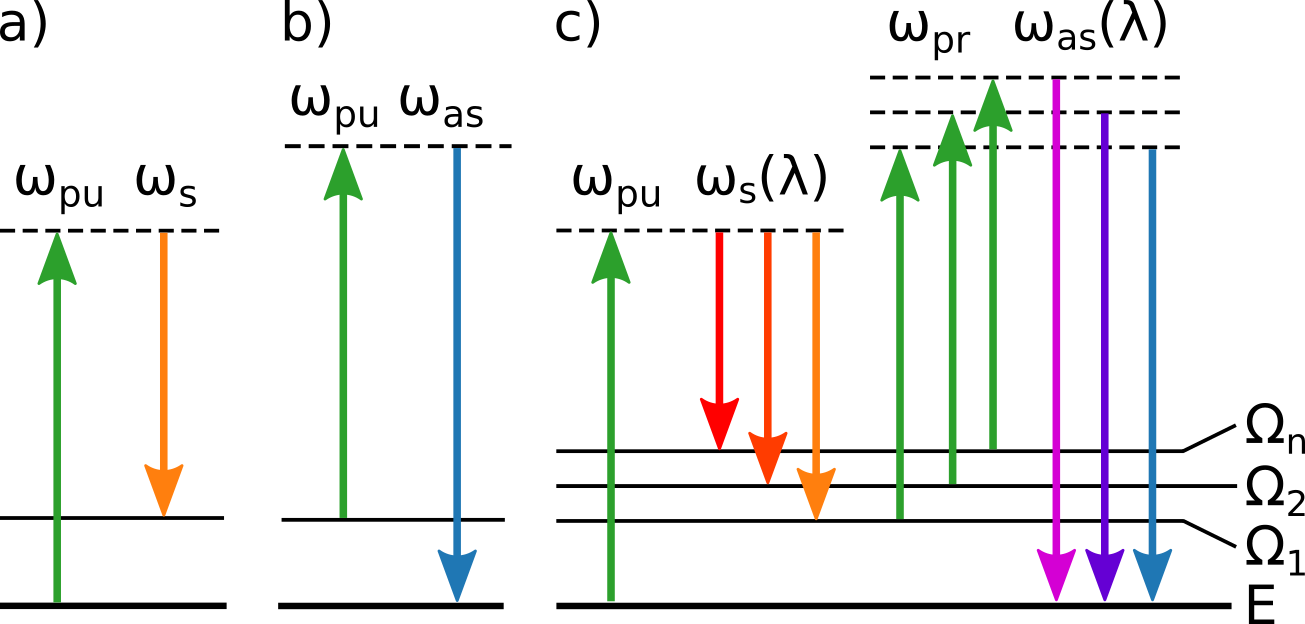}
	\caption[Raman scattering Jablonski Diagram]{Jablonski diagrams of a) spontaneous Stokes scattering, b) spontaneous anti-Stokes scattering, and c) BCARS.
\label{fig:jablonski}}
\end{figure}

The basis of Raman spectroscopy is inelastic light-matter scattering. For SR, a pump photon of frequency $ \omega_{pu} $ scatters inelastically to either lose (Stokes scattering, figure \ref{fig:jablonski}.a) or gain energy (anti-Stokes scattering, figure \ref{fig:jablonski}.b) corresponding to the probed vibrational level $ \Omega $.
The scattering cross-section for SR is low due to its spontaneous and incoherent nature.
In contrast, BCARS is a nonlinear process where the sample interacts with three fields
and the vibrational levels $\Omega_i$ are populated through coherent excitation at $ \omega_{pu} $ and $ \omega_s $, whose energy difference corresponds to the vibrational level. These states are read out by the third field $ \omega_{pr} $, resulting in the emission of an anti-Stokes field $ \omega_{as} $. By using a broadband light source for $ \omega_s $, a broad range of vibrational modes can be probed, leading to the generation of a broadband anti-Stokes component (figure \ref{fig:jablonski}.c).

\subsection{Extraction of Spectral Information}

The measurable BCARS signal $ I_{CARS} $ depends on the spectral response of the sample and the detector response function $ \tilde{C}(\omega) $
as follows\cite{Camp2016}:

\begin{align}
    I_{CARS}(\omega) \propto |\tilde{C}(\omega)|^2 |\chi^{(3)}(\omega)|^2 \label{eq:I_cars}
\end{align}

Here, $ \chi^{(3)} $ describes an effective susceptibility containing convolution of the actual susceptibility and the probe pulse.
As BCARS is a third-order nonlinear optical process, the spectral response of the sample is described by its third-order susceptibility tensor $ \chi^{(3)}(\omega) $\cite{lotem1976}, which consists of two terms, a resonant and a non-resonant one:

\begin{align}
    \chi^{(3)} &= \chi^{(3)}_R + \chi^{(3)}_{NR} \label{eq:chi3}
\end{align}

The resonant part $ \chi^{(3)}_R $ is the sum of many complex Lorentzian functions, and its imaginary part corresponds to the SR-like response of the sample.
The laser excitation stimulates an additional chemically unspecific background of electronic signals, the NRB.
It corresponds to the non-resonant $ \chi^{(3)}_{NR} $, which can be considered purely real under the assumption that the pump/Stokes frequencies are far from electronic resonances.
Therefore, extracting the imaginary part of the resonant term from the BCARS intensity is necessary to obtain SR-like spectral features.
In this work, we remove the NRB from the BCARS spectra by adapting the Python-based algorithm Crikit2 developed by Camp et al. \cite{Camp2016}, based on the time-domain KK transformation. The detailed mathematical description can be found elsewhere \cite{Camp2016}.

\section{Methods}

For our BCARS measurements, we employed two different experimental configurations: The first one is the CERES setup in Dresden, Germany, which is a commercial setup optimized for the investigation of crystalline materials and their lower-shift Raman modes ($\nu \approx \SI{200}{}-\SI{1100}{\per\cm}$). The second system is the tailor-made VIBRA setup in Milan, Italy, which is optimized for high-speed imaging across the entire fingerprint region ($\nu \approx \SI{500}{}-\SI{1800}{\per\cm}$). Another key difference is the detection geometry: While CERES operates in epi-detection, VIBRA can measure in both transmission- and epi-direction.

\subsection{CERES Setup}

\begin{figure}[htbp]
	\includegraphics[width=0.4\textwidth]{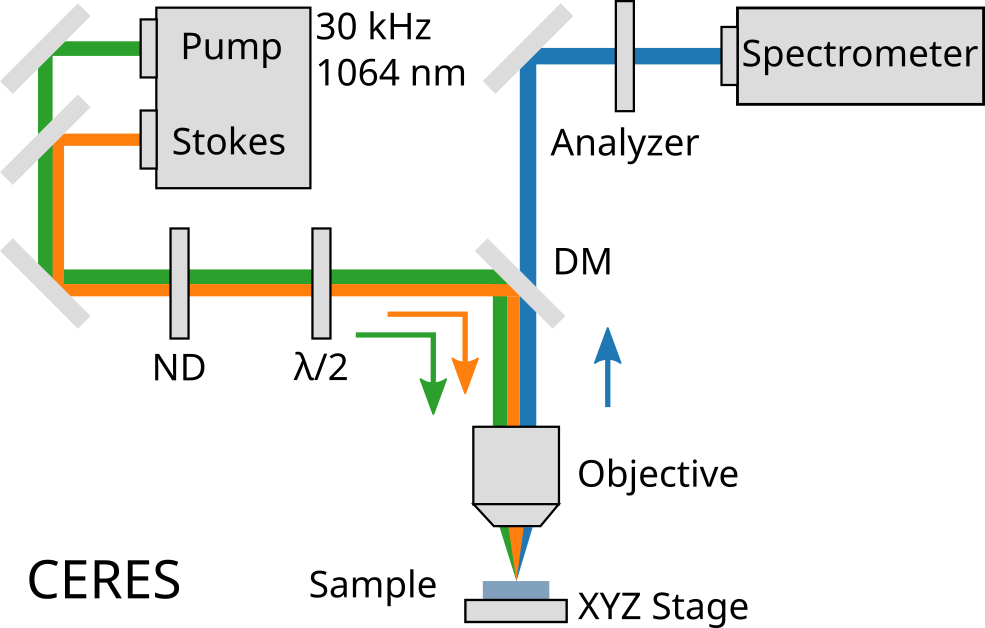}
	\caption[CERES setup]{
	Layout of the CERES BCARS setup used in this work: The pulsed incident pump (\SI{1064}{\nm}) and broadband Stokes laser (\SI{600}{}-\SI{2000}{\nm}) are spatially and temporally matched. Measurements are conducted in epi-detection geometry. ND: neutral density filter, $ \lambda/2 $: rotatable half-wave plate, DM: dichroic mirror.
	\label{fig:ceres_setup}}
\end{figure}

The CERES setup consists of a LabRAM HR Evolution commercial Raman microscope (HORIBA Jobin Yvon GmbH, Oberursel, Germany), as shown in figure \ref{fig:ceres_setup}.

The pump and Stokes pulses are generated by a commercial combined laser system (LEUKOS CARS-SM-30), which consists of a Q-switched microchip Nd:YAG laser in LP$_{01}$ single-mode configuration pumping a photonic crystal fiber to generate a broadband supercontinuum.
The system produces \SI{1064}{\nm}, \SI{100}{\mW} average power pump pulses and \SI{600}{}-\SI{2000}{\nm}, \SI{80}{\mW} average power Stokes pulses, both with a \SI{1}{\ns} pulse length and a \SI{30}{\kilo\hertz} repetition rate.
The Stokes pulses are used in the range of \SI{1085}{}-\SI{2000}{\nm} covering the BCARS signal range of \SI{190}{}-\SI{4000}{\per \cm}.
The optical paths are adjusted to ensure the pulses' temporal and spatial overlap on the sample.
SR measurements can be performed alongside, using a \SI{632.8}{\nm} monochromatic He-Ne continuous wave laser (Melles Griot) as an alternative laser source.
A rotatable half-wave plate controls the laser's polarization.
A Nikon CFI APO NIR 40$\times$ water-immersion objective ($NA = 0.8$) focuses the pulses on the sample and collects the scattered light in epi-detection.
A dichroic mirror acts as the short-pass filter and transmits the blue-shifted BCARS signal to the detector, which consists of a spectrometer with a \SI{600}{lines\per \mm} diffraction grating, and a Syncerity CCD detection system (HORIBA Jobin Yvon GmbH, Oberursel, Germany), resulting in a spectral resolution of the detector of \SI{0.05}{\nm} (\SI{0.45}{}-\SI{0.56}{\per\cm}) in the relevant spectral range.

The sample is mounted on an XYZ translation stage, which enables raster scanning in three dimensions.

\subsection{VIBRA Setup}

\begin{figure}[htbp]
	\includegraphics[width=0.45\textwidth]{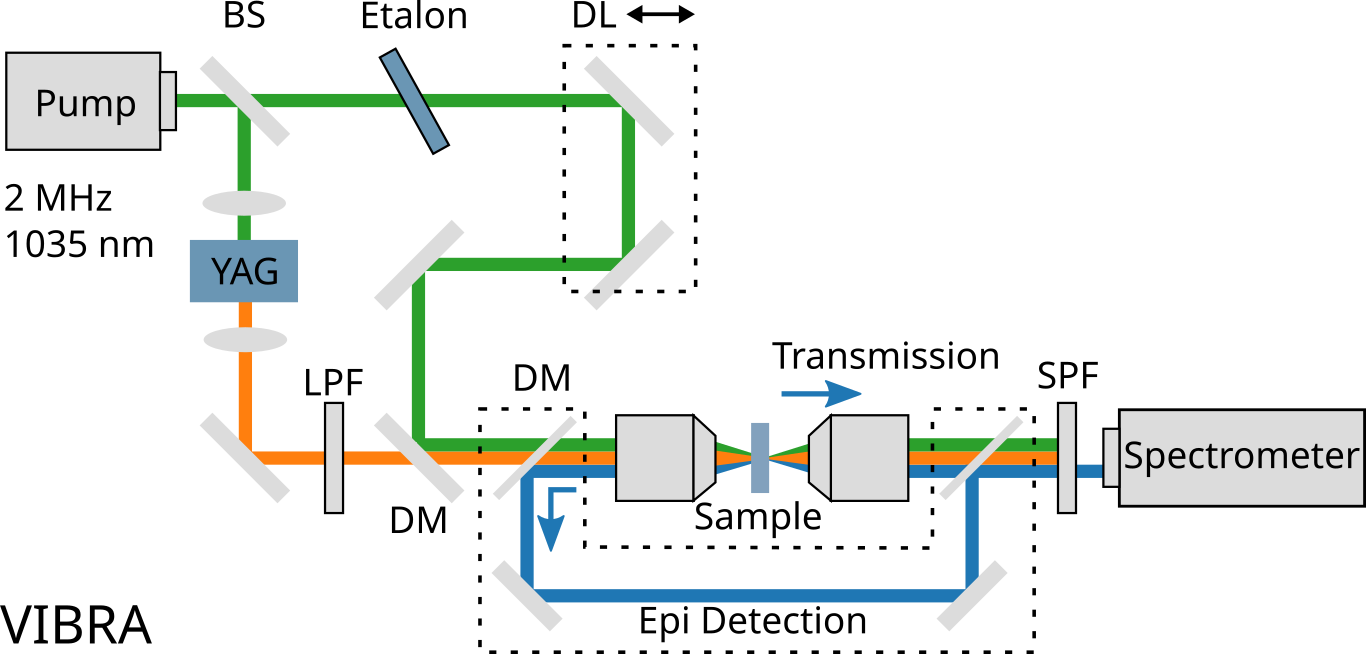}
	\caption[VIBRA setup]{
	Layout of the VIBRA BCARS setup used in this work: The incident laser source is a \SI{1030}{\nm}, \SI{2}{MHz} pulsed laser, which is split by a beam splitter (BS): The first part is spatially broadened by a YAG crystal and a long pass filter (LPF) selects the red-shifted part. The second part is spectrally filtered by an etalon and temporally shifted by a delay-line (DL) to ensure temporal and spatial overlap at the sample. The BCARS signal can be detected either in transmission or in epi-detection geometry. DM: dichroic mirror, SPF: short-pass filter.
	\label{fig:monaco_setup}}
\end{figure}

The architecture of the VIBRA BCARS microscope is shown in figure \ref{fig:monaco_setup} and described in detail elsewhere \cite{Vernuccio:22}.

A commercial fiber-based ytterbium laser system (Coherent Monaco) provides \SI{270}{\femto\second} pulses at a \SI{1035}{\nm} wavelength with a \SI{2}{\mega \hertz} repetition rate. A beam splitter divides the laser output into two beams.
The first beam generates narrowband pump pulses through a high-finesse Fabry-Perot etalon to decrease the pulse bandwidth.
This determines the spectral resolution of \SI{10}{\per\cm} (\SI{1.1}{\nano \meter} full width at half maximum (FWHM) bandwidth).
The second replica generates a broadband Stokes pulse via white-light continuum generation by focusing into a \SI{10}{\mm}-thick yttrium aluminum garnet (YAG) crystal \cite{Grazuleviciute2015}.
A long-wave-pass filter selects the red-shifted lobe of the Stokes pulse (\SI{1050}{}–\SI{1300}{\nm}). In front of the microscope, the pump and Stokes pulses are collinearly superimposed through a dichroic mirror and temporally synchronized by a mechanical delay line.
The beams are then sent to a homebuilt transmission microscope in up-right configuration, equipped with two identical 100x air objectives of $NA = 0.85$ (Olympus LCPLN100XIR).
Behind the sample, a short-pass filter (FESH1000, Thorlabs) rejects the pump and Stokes beams, transmitting the generated BCARS signal, whose spectrum is then measured with a standard grating-based dispersive spectrometer (ACTON SP-2150, Princeton Instruments) using a CCD detector (Blaze 100HR, Princeton Instruments).
Additionally, a movable dichroic mirror (DMLP1000, Thorlabs) is inserted in front of the illuminating objective, allowing to perform BCARS detection in epi-configuration.
The sample is raster-scanned in three dimensions using a motorized XYZ translation stage synchronized with the CCD camera of the spectrometer.

\subsection{Samples}

For our internal Round Robin, crystalline materials constitute ideal reference samples.
They are easily transportable and stable under environmental changes like temperature, pressure, and humidity, which might differ between laboratories.
Moreover, the sample preparation is reproducible, samples do not degrade, and they generally have sharp Raman peaks, which is ideal for measuring minor changes in their spectra.
We have chosen four materials with varying structural complexity to increase the number of detectable Raman modes and to show the variety of possible applications.

The complexity of the crystals surveyed in this paper is gauged using a complexity measure introduced by Krivovichev  \cite{Krivovichev2014} and later expanded by Hornfeck. The open-source Python script crystIT by Kaussler and Kieslich \cite{Kaussler:oc5005} is used to calculate the complexity
to determine the amount of information in a crystal unit cell.

\paragraph{Diamond} Diamond is a cubic crystal of carbon with space group Fd$\Bar{3}$m \cite{Villars2016:sm_isp_sd_1500919}. Its unit cell contains eight carbon atoms. However, its configurational complexity per unit cell is zero because it lacks any coordinational degrees of freedom. It has one fundamental Raman-active $ F_{2g} $ mode \cite{Krishnan1944}. The diamond used in this study is a polished single-crystal CVD diamond cut along the [100] orientation with a thickness of \SI{200}{\micro m} manufactured by Applied Diamond, Wilmington (DE), USA.

\paragraph{6\ch{H-SiC}} The 6H polytype of SiC used in this paper is of hexagonal structure and space group P6$_3$mc \cite{Villars2016:sm_isp_sd_1628877}, with eight atoms in its unit cell, resulting in a configurational complexity of 35. Expected Raman modes are $3 A_1$, $3 E_1$, and $4 E_2$ \cite{Feldman1968}. The sample is a z-cut 6H-SiC from MSE Supplies, Tucson (AZ), USA.

\paragraph{\ch{D2K[PO4]}} (KDP) is tetragonal with a space group I$\Bar{4}$2d \cite{Villars2016:sm_isp_sd_1145237} and 32 atoms per unit cell, resulting in a configurational complexity of 68.7. Theroretically expected Raman modes are 11 $A_1$, 13 $B_1$, 12 $B_2$, and 23 $E$ \cite{Lu2002}. The sample used here is a z-cut crystal provided by the Czech Academy of Sciences, Prague, Czech Republic.

\paragraph{\ch{KTiOPO4}} (KTP) is orthorhombic with space group Pna2$_1$ \cite{Allan1996} and 64 atoms in its unit cell. Its configurational complexity is thus 558.3, constituting a complex ion conductor with a vibrant Raman spectrum with an astonishing 47 $A_1$, 47 $B_1$, 47 $B_2$, and 48 $A_1$ modes. The sample used here is a z-cut KTP provided by the Czech Academy of Sciences, Prague, Czech Republic.

\section{Results}

\subsection{Phase Retrieval}

\begin{figure}[htbp]
	\includegraphics[width=0.45\textwidth]{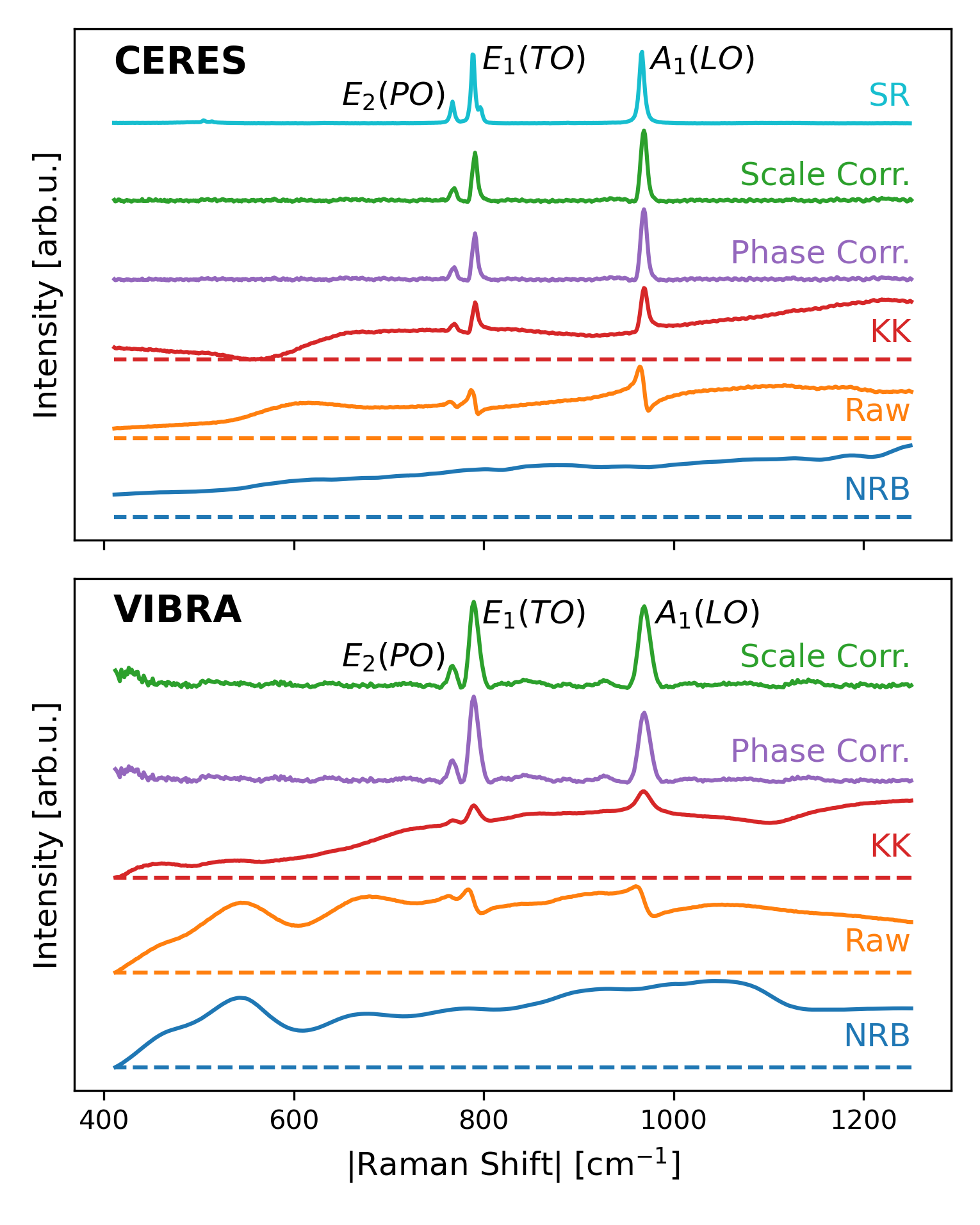}
	\caption[KK-Transformation Steps]{
	Transformation steps to retrieve the SR-like spectrum using the CRIKit2 algorithm demonstrated with the example of 6H-SiC for both the CERES and VIBRA setup: The NRB (blue) is measured on glass and the raw, epi-detected BCARS signal (orange) on 6H-SiC. The exposure times are \SI{500}{\ms} for CERES, and \SI{10}{\ms} for VIBRA. The Kramers-Kronig transformed signal (red) is furthermore corrected for phase errors (purple) and scaling errors (green). Finally, the SR spectrum (cyan) was measured in CERES (\SI{15}{\second} exposure time) for comparison. The dashed lines indicate zero-intensity levels.
	\label{fig:trafo_steps}}
\end{figure}

Comparing the raw BCARS spectra between laboratories is challenging, as the distortion by the NRB is setup-dependent: The NRB shape is substantially influenced by the spectral profile of the broadband Stokes pulse, the pump wavelength, the detector response function, and the spectral profile of the diffraction grating. Additionally, each optical component within the laser path has a wavelength-dependent influence on polarization and intensity.
In order to compare the results and estimate the reproducibility of BCARS for the analysis of crystalline materials, one needs to use phase-retrieval algorithms to properly remove the NRB response and extrapolate the purely resonant contribution of the sample, independent of the power spectral density of the sources.

Therefore, the first step is the removal of the NRB, for which multiple approaches are available.
Deep-learning techniques are shown to be most reliable \cite{Houhou2020} but require a set of training data of similar spectra. In our case, using data from two different setups and different NRB shapes would have required further efforts for the generation of a suitable training dataset (variable number of spectral pixels, spectral shape, amplitude and width of the peaks) to obtain accurate prediction which is out of the scope of the current work.
Here, the KK transformation is chosen, which is shown to be equivalent to the maximum entropy method \cite{Cicerone2012}.

The multiple transformation steps are shown exemplarily for 6H-SiC in figure~\ref{fig:trafo_steps}.
Note that BCARS detects the blue-shifted anti-Stokes component, exhibiting negative Raman-shift values. For comparison to SR data and easier readability, we give the absolute values for Raman shifts and peak positions, i.e., $|$Raman Shift$|$. The Raman shift axes for both setups are calibrated using emission lines from the same Neon-gas lamp (Model 6032, Newport Spectra-Physics GmbH, Darmstadt, Germany).

The pure NRB was measured by taking the BCARS spectrum of borosilicate glass, which shows no distinct Raman peaks. Note that this is not the correct NRB of the investigated 6H-SiC, as it is impossible to measure only the non-resonant part of the signal inside a crystal without recording the resonant response of the sample as well. Both raw and NRB spectra were corrected by subtraction of a dark spectrum.
The BCARS spectra (orange) differ between the two setups but have a broad and almost constant shape in both cases. An exception is the VIBRA spectrum at around \SI{500}{\per\cm}, where a broad intensity peak is measured, mainly due to self-phase modulation occurring during the supercontinuum generation in the YAG crystal.
The measured single-point spectra show a high signal-to-noise ratio and do not require further processing through denoising methods contained in CRIKit2.
The phase retrieval is done with the KK transformation in CRIKit2 using the default parameters.
The retrieved spectra (KK, red) exhibit Lorentzian-shaped peaks and a varying baseline due to using the NRB on glass as a reference instead of the inaccessible "real" NRB.
CRIkit2 runs through two steps of error correction to remove the baseline in the complex phase of the signal (Phase Corr., purple) and correct the relative intensities of the peaks (Scale Corr., green).
For both steps, the parameters are adjusted individually for CERES and VIBRA.
For the precise values used for all transformation processes, see the Supplemental Material at [URL will be inserted by publisher] .
The scale error corrected spectrum (green) is closest to an SR-like spectrum and, in the following, will be used to discuss the phase-retrieved (PR) spectrum.

\begin{table}[htbp]
         \caption[SiC Phonon Assignment]{Main peak frequencies $\Delta\tilde{\nu}_{Peak}$ detected in BCARS and SR measurements on SiC, and phonon assignment based on reported phonon frequencies $\Delta\tilde{\nu}_{Phonon}$ by SR \cite{Burton1998}.
        \label{tab:SiC}}
 \begin{ruledtabular}
 \begin{tabular}{ |c c c|c|c| }
\hline
\multicolumn{3}{|c|}{$\Delta\tilde{\nu}_{Peak}$ [cm$^{-1}$]} & Assigned & $\Delta\tilde{\nu}_{Phonon}$  [cm$^{-1}$] \\
VIBRA & \multicolumn{2}{c|}{CERES} & Phonon & (SR) \cite{Burton1998} \\
BCARS & BCARS & SR & &  \\
\hline
767 & 768 & 767 & $E_2$(PO)  & 767.5 \\
790 & 791 & 789 & $E_1$(TO) & 788.0 \\
969 & 969 & 966 &  $A_1$(LO) & 966.5 \\

\hline
\end{tabular}
\end{ruledtabular}
\end{table}

After the phase retrieval processing, the spectra from both setups have Lorentzian-shaped, SR-like peaks, as reported in table \ref{tab:SiC}. CERES could detect modes at \SI{768}{\per\cm} ($E_2 PO$), \SI{791}{\per\cm} ($E_1 TO$) and \SI{969}{\per\cm} ($A_1 LO$), with different relative intensities.
VIBRA measures the peaks at \SI{767}{\per\cm}, \SI{790}{\per\cm} and \SI{969}{\per\cm}, respectively, which are, within error ranges, located at the same positions.
One should note that the retrieved peaks for the VIBRA setup are broader because of the broader bandwidth of the pump pulse (\SI{10}{\per\cm} vs. \SI{0.6}{\per\cm} in the CERES setup).
The absolute signal intensities can best be compared for the raw spectra, where the resulting relative peak intensities differ: While VIBRA has almost equal intensities for the \SI{791}{\per\cm} and \SI{969}{\per\cm} modes, CERES detects a slightly more intense signal at \SI{969}{\per\cm}.
These results show that comparable spectra can be retrieved despite the vastly different raw and NRB data. Furthermore, the Raman shift axis calibration ensures the comparability of the peak positions. Finally, the BCARS results align well with SR measurements and theory, proving the applicability of the technique to solid-state materials independent of setups.


\subsection{Epi- vs. Transmission Detection}

For transparent materials or thin films, it is possible to detect the signal either in the forward (transmission) or backward (epi) direction.
However, the signal strength for these directions differs strongly due to the wave-vector missmatch $\Delta k$ and the corresponding coherent interaction length $l_c = \pi / \Delta k$.
The $l_c$ dictates the range over which the signal is constructively accumulated.
In the forward scattered signal, $\Delta k = 2 k_{pu} - k_s - k_{as}$ is close to zero, resulting in an $l_c$ on the micrometer length scale, which is longer than the confocal parameter of the focused pump/Stokes beams.
Whereas for epi-scattering with a backward propagating anti-Stokes beam, there is a significant k-vector mismatch $\Delta k = 2 k_{pu} - k_s + k_{as}$ resulting in $l_c$ being a fraction of the anti-Stokes wavelength \cite{Cheng2004}.

\begin{figure*}[htbp]
	\includegraphics[width=0.90\textwidth]{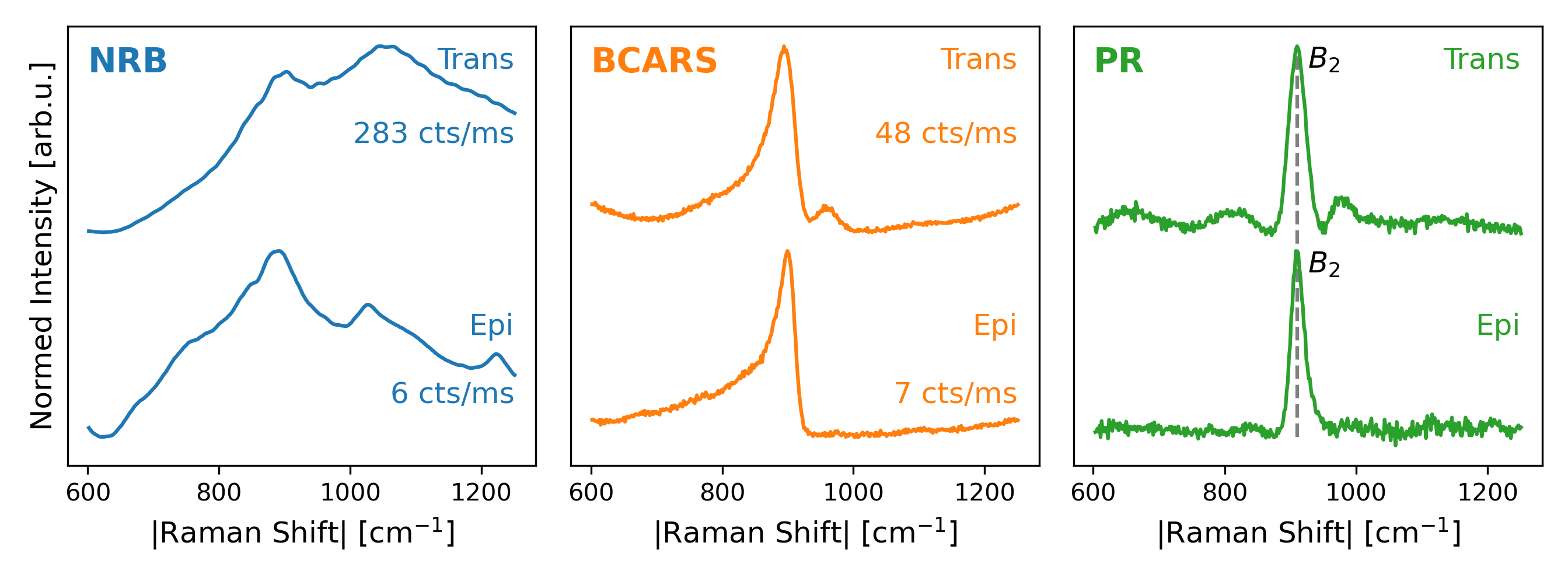}
	\caption[Epi- vs. transmission detection]{
	Comparison of the transmission- (trans, upper row) and epi-detected signal (epi, lower row) of KDP using the VIBRA setup: Both the NRB (blue) and the raw BCARS data (orange) show almost the same spectral shape, although in transmission the intensity is \SI{6}-\SI{47} times higher. The phase-retrieved (PR) data (green) shows the same spectral features, thus rendering both signals equivalent in information content.
	\label{fig:epi_trans}}
\end{figure*}

The epi signal is used when imaging heterogeneous biological samples such as cells and tissues since they feature scatterers whose dimension is much smaller than $l_c$ \cite{Cheng2004}.
Using a high-NA objective provides a range of k-vectors to always fulfill the phase-matching condition \cite{potma2000, Amber2021, Amber2022, Spychala2023}.
However, for bulk crystalline samples, the scattering direction is consequential, and previous results in crystals concluded that there is no measurable epi-scattered signal. Instead, the back-detected signal is the strong forward signal reflected on the backside of the sample \cite{Hempel2021}.
This hypothesis is tested by measuring the BCARS spectra of bulk KDP and measuring the NRB spectra of a borosilicate glass slide in both directions at the same position near the backside of the sample.
These experiments are done in the VIBRA setup by introducing an additional dichroic mirror in front of the illumination objective to collect the back reflected light and send it to the spectrometer, thus switching between transmission- and epi-detection. The results are shown in figure \ref{fig:epi_trans}.

The measured spectra for NRB (blue) and raw BCARS (orange) are similar in shape, except for some broad features in the NRB epi spectrum at \SI{1050}{\per\cm} and \SI{1250}{\per\cm}, and in the raw transmission spectrum at \SI{900}{\per\cm} due to modulation of the broadband Stokes beam.
The similarity is expected, as the same laser source and detector are used.
However, the differences in spectral shape can be explained by the different optical components present along the two detection paths. For both cases, the transmission signal is stronger - by a factor of \SI{47}{} and \SI{6.8}{} for the NRB and the BCARS signal, respectively. The ratio between transmission- and epi-detected signal strength might vary between samples, depending on their thickness, transparency, and surface roughness.
The retrieved spectra (green) contain equivalent spectral information: The $B_2$ mode is measured at \SI{911}{\per\cm} and \SI{910}{\per\cm} for transmission- and epi-detection, respectively.
This difference is smaller than the spectral resolution of the setup.
The transformation algorithm delivers similar results when given similar input spectra.
The presence of spurious modulations in the transmission spectrum is due to self-phase modulation processes occurring during the generation of the white light in the YAG crystal.
Further expected peaks from KDP cannot be measured as their Raman shifts are smaller than the detection limit of \SI{500}{\per\cm} of the VIBRA setup.
In conclusion, the forward- and backward-detected signals show the same Raman peaks. However, the forward scattered signal should be measured to achieve higher signal intensity if the sample and setup allow it.

\subsubsection*{Focal Depth Dependence}

\begin{figure}[htbp]
	\includegraphics[width=0.45\textwidth]{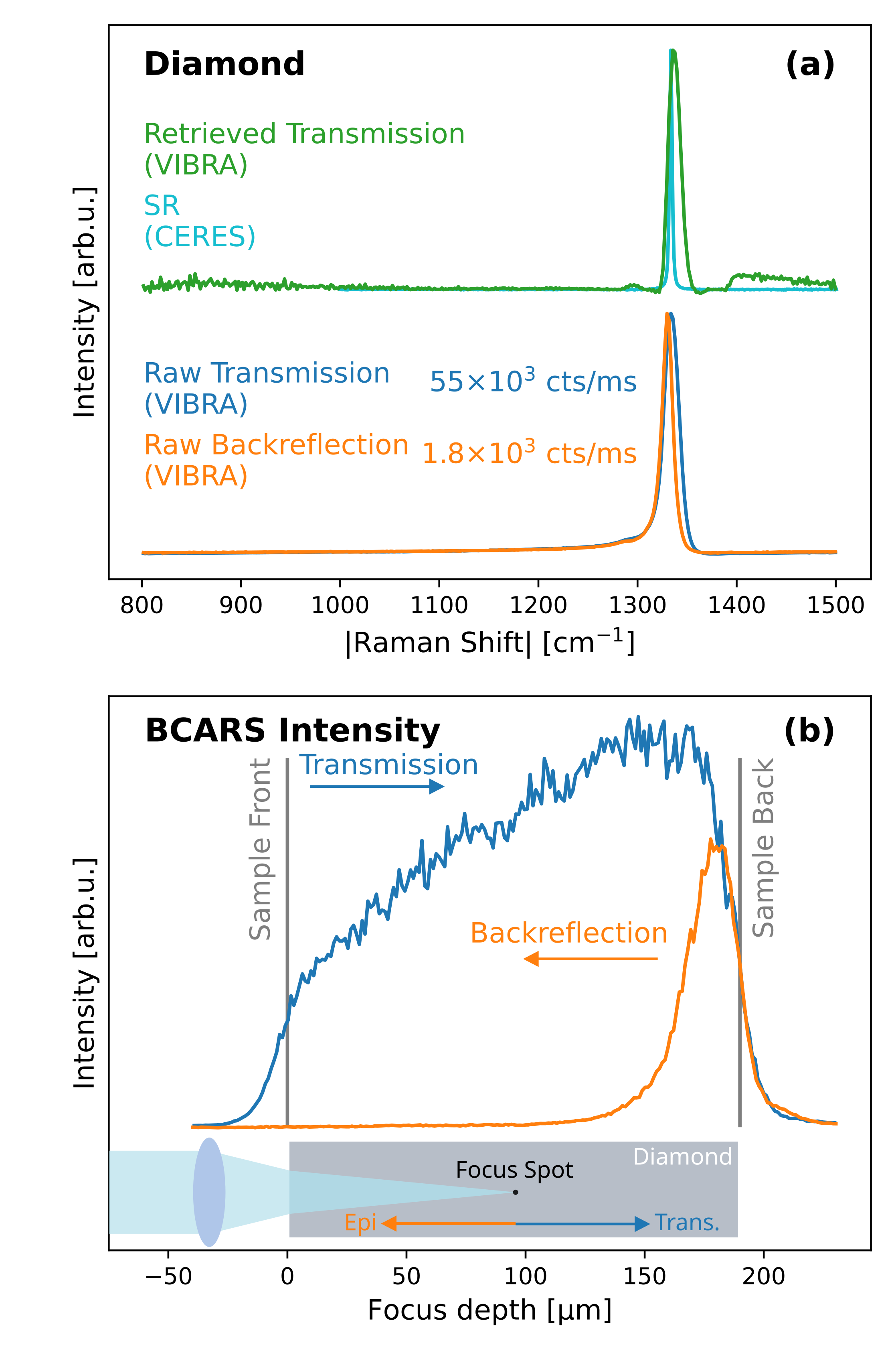}
	\caption[Depth Scans]{BCARS on a diamond single crystal measured with VIBRA. a) Raw BCARS spectra in transmission (blue) and epi-detection (orange) yield the same results after phase retrieval processing (green), which is compared to the SR spectrum (cyan) measured with CERES. b) Focal depth scan through diamond single crystal: In transmission detection (blue), the integrated signal intensity is measurable along the whole crystal, from the front- to the backside. On the other hand, the epi-detected signal (orange) is only detectable near the backside, where the strong forward scattered signal is reflected.
	\label{fig:z-stack}}
\end{figure}

When measuring bulk samples significantly thicker than the laser Rayleigh range, the dependence of the focal spot position inside the crystal is a fundamental parameter to consider. Therefore, we performed depth scans on a diamond single crystal to compare the BCARS signal in forward and epi-detection depending on the focal position, as shown in figure \ref{fig:z-stack}.

Diamond was shown to have an intense CARS peak at the $sp^3$ vibrational resonance \cite{Pope2014}. The spectra from VIBRA shown in figure \ref{fig:z-stack} (a) show a single strong peak at \SI{1336}{\per\cm}. It has an FWHM of \SI{14}{\per\cm}, with good agreement to the SR comparison measurements at \SI{1333}{\per \cm} (FWHM = \SI{3.1}{\per\cm}) in CERES and the literature value of \SI{1332}{\per\cm} (FWHM down to \SI{1.7}{\per\cm}) \cite{Knight1989}.
With signal intensities of \SI{55e3}{cts\per\ms} for transmission and \SI{1.8e3}{cts\per\ms} for the epi-detected signal, this is two orders of magnitude stronger than any other material measured in this work.
Figure \ref{fig:z-stack} (b) shows the BCARS intensity upon longitudinal translation of the focal spot inside the material.
The transmission signal shows a linear increase along the depth axis due to lower scattering and losses of the BCARS signal when focusing closer to the backside of the sample.
The signal in epi-detection shows a strong intensity near the backside of the sample, indicating that the reflection of the forward signal on the far surface is measured.
A backscattered signal would show a rather constant intensity across the sample, with a decrease further inside due to scattering losses.
As the same objective is used for focusing the laser and collecting the signal, the best collection efficiency is achieved whenever the reflection point at the backside is in focus and the signal beam is not widened.
Therefore, it is advised to measure epi-BCARS near the backside of the sample.

The signal intensity changes drastically at the surfaces as the laser focus volume enters or leaves the signal-generating material.
The spectral information is independent of focus depth in a single crystalline material, except for possible surface phenomena. On the other hand, for quantitatively comparable measurements, measuring in the same depth and having the same material thicknesses is essential.

\subsection{Setup Comparison}

\begin{figure*}[htbp]
	\includegraphics[width=0.9\textwidth]{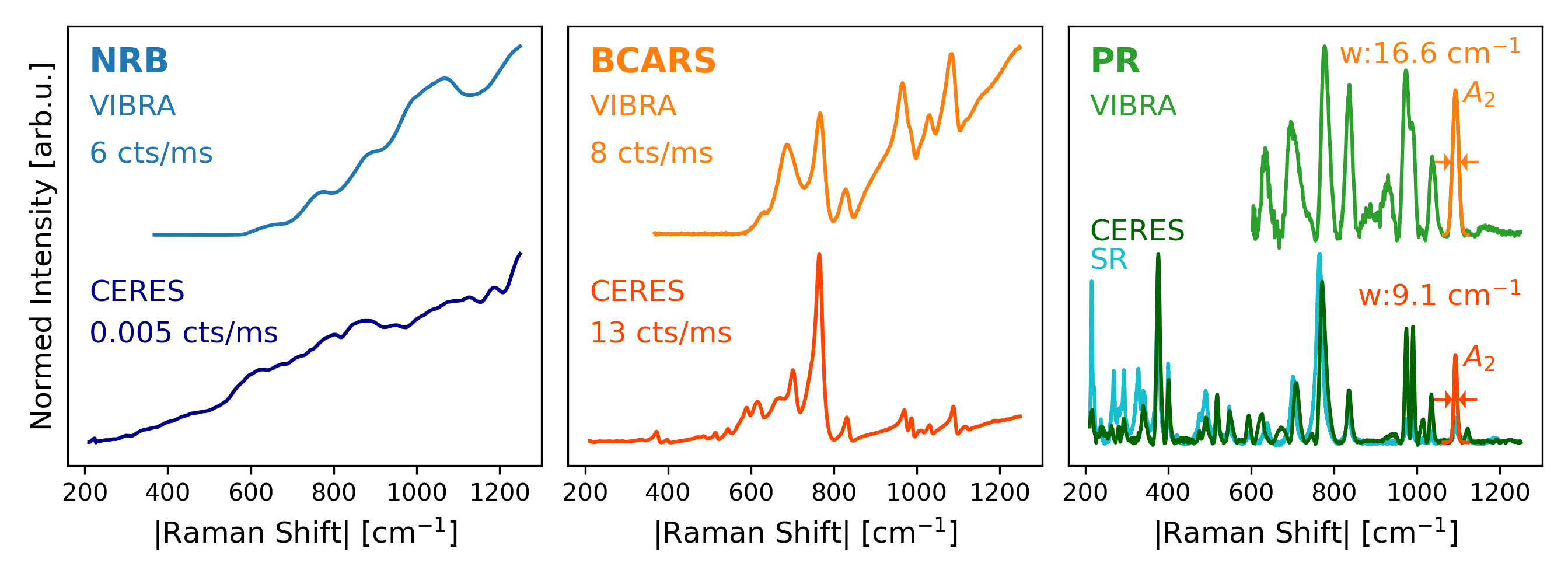}
	\caption[Setup comparison on KTP]{
    Comparison of BCARS signals from the same KTP sample measured in the two different setups: VIBRA (upper row) and CERES (lower row). The NRB (blue) shows different shapes, which influence the baselines of the raw BCARS spectra (orange). The distorted KTP peaks can be seen at roughly the same position but with vastly different shapes and intensities for the two setups. The PR spectra (green) compare to the SR result (cyan). Again, the spectra show matching spectral positions but differences in relative intensities and width.
	\label{fig:setup_comparison}}
\end{figure*}

Reproducibility and comparability between different setups are essential aspects of scientific research. As stated before, this is especially important for BCARS, as the NRB signal strongly depends on various setup parameters such as laser source, focusing, and detector response function.
As a comparison study, we have measured the BCARS spectra of the same KTP sample in both the CERES and VIBRA setups using epi-detection. The results are shown together with the NRB data from glass in figure \ref{fig:setup_comparison}.

The NRB spectra show a similar baseline with decreasing intensity for decreasing wavenumbers. However, the shape differs due to different Stokes laser profiles and detector response functions for both setups.
The raw BCARS spectra show peaks at a similar position and with BCARS-typical peak deformation, but they differ in baseline shape and peak widths.
Additionally, CERES has a larger detection range, as its small bandwidth filter allows for detection down to \SI{190}{\per\cm}, while the VIBRA data is cut at \SI{500}{\per\cm}.
The signal intensity for epi-detected raw BCARS is on the same scale for both setups.
For comparison, the VIBRA transmission detection yields a \SI{E3}{} times higher signal strength. On the other hand, for the NRB, VIBRA achieves a \SI{E3}{} times greater intensity in counts per second because a thinner glass slide was used.
After phase retrieval processing, both spectra show Lorentzian-shaped peaks and their positions align with the corresponding SR spectrum measured in CERES.
However, the relative peak intensities differ for both BCARS and SR spectra.
The fitted peak positions and their mode assignment are described in table \ref{tab:KTP}.

\begin{table}
         \caption[KTP Phonon Assignment]{Main peak frequencies $\Delta\tilde{\nu}_{Peak}$ detected in BCARS and SR on KTP, and phonon assignment based on reported phonon frequencies $\Delta\tilde{\nu}_{Phonon}$.
        \label{tab:KTP}}
 \begin{ruledtabular}
 \begin{tabular}{ |c c c|c|c| }
\hline
\multicolumn{3}{|c|}{$\Delta\tilde{\nu}_{Peak}$ [cm$^{-1}$]} & Assigned & $\Delta\tilde{\nu}_{Phonon}$  [cm$^{-1}$] \\
VIBRA & \multicolumn{2}{c|}{CERES} & Phonon & (SR) \cite{Kugel1988, Vivekanandan1997} \\
BCARS & BCARS & SR & &  \\
\hline
- & 216 & 215 & $A_1$ & 213 \\
- & 262 & 268 & $A_1$ & 268 \\
- & 280 & 280 & $A_2$ & 284 \\
- & 293 & 293 & $B_2$ & 304 \\
- & 340 & 341 & $A_2$ & 335 \\ 
- & 376 & 374 & $A_1$ & 378 \\
- & 400 & 400 & $\nu_2$\ch{PO4} & 400 \\
- & 491 & 490 & $\nu_4$\ch{TiO6} & 509 \\
- & 518 & 517 & $A_1$ & 517 \\
- & 549 & 547 & $\nu_4$\ch{PO4} & 548 \\
- & 594 & 591 & $A_1$ & 598 \\
634 & 624 & 636 & $\nu_1$\ch{TiO6} & 632 \\ 
693 & 673 &  & $B_2$ & 673 (IR) \\
709 & 709 & 701 & $\nu_2$\ch{TiO6} & 709 \\
779 & 773 & 764 & $A_1$ & 770  \\ 
836 & 836 & 835 & $\nu_3$\ch{TiO6} & 833 \\
975 & 974 & 975 & $\nu_1$\ch{PO4} & 974 \\
990 & 991 & 990 & $A_1$ & 993 \\
 & 1013 & 1013 & $A_1$ & 1008 \\
1037 & 1034 & 1034 & $A_1$ & 1034 \\
1093 & 1093 & 1093 & $\nu_3$\ch{PO4} & 1095 \\
1171 & 1122 & 1120 & $A_1$ & 1113 \\

\hline
 \end{tabular}
 \end{ruledtabular}
 \end{table}

One notable difference is in the peak width, exemplarily shown for the \SI{1093}{\per\cm} $A_2$ mode: CERES detects an FWHM of \SI{9.1}{\per\cm} in BCARS and \SI{6.7}{\per\cm} for SR.
Using a higher-resolution diffraction grating does not decrease the FWHM of the mode, which indicates that the detector resolution does not limit its width.
In comparison, VIBRA shows an FWHM of \SI{16.6}{\per\cm}. Due to the broader width, peaks can overlap, making the distinction between close modes challenging, as seen for the \SI{836}{\per\cm} $A_1$ mode.
The detector resolutions were tested by measuring a Neon gas lamp spectrum: CERES has a minimum peak width of \SI{0.25}{\nm} or \SI{2.8}{}-\SI{3.9}{\per\cm}, which is three times smaller than the measured peak width.
As the spectral width of CERES' \SI{1}{\nano\second} laser pulse corresponds to a \SI{0.015}{\per\cm} spectral resolution, it can be assumed that the measured peak width corresponds to the actual width of this KTP mode.
In contrast, the VIBRA detector has a minimum peak width of \SI{1.2}{\nm}, or \SI{13.8}{\per\cm}. 
Here, the spectral resolution is limited by the short laser pulses to $\approx \SI{10}{\per\cm}$.
However, the shorter laser pulses have a higher maximum intensity, which allows for a higher BCARS signal intensity and lower pixel dwell time.
Additionally, the spectral resolution of VIBRA can be increased, if needed, by changing the etalon to decrease the spectral width of the pump signal.
In conclusion, both setups can detect equal spectral information. Depending on the objectives of the conducted experiment, the setup can be optimized for higher spectral resolution or faster acquisition by varying the pump and Stokes pulse parameter.

\subsection{Imaging}

\begin{figure*}[htbp]
	\includegraphics[width=0.9\textwidth]{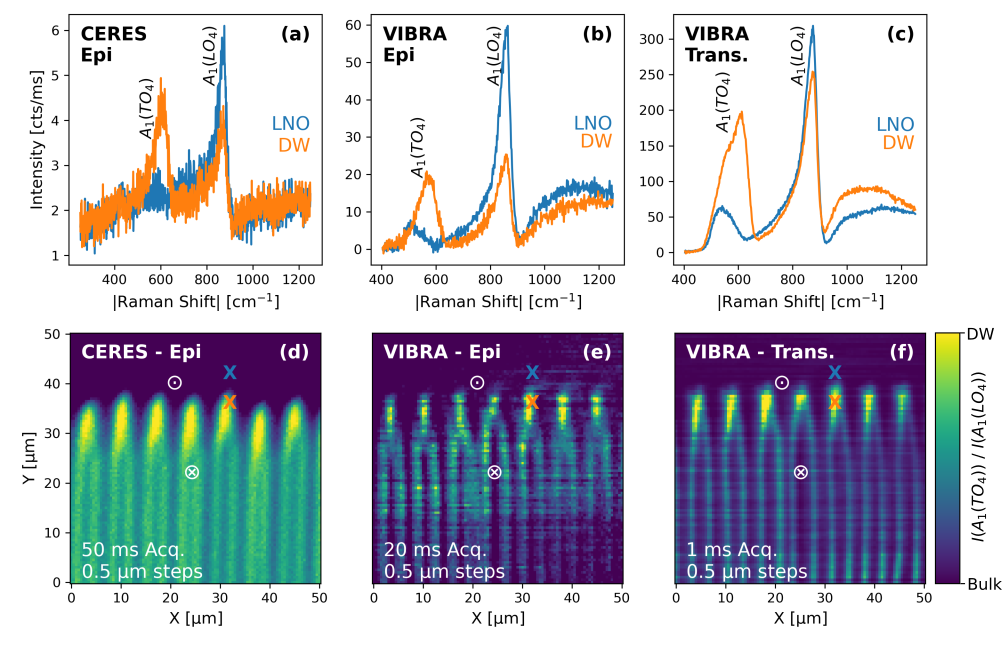}
	\caption[Domain Wall Imaging]{
    BCARS  imaging of domain walls in periodically poled Lithium Niobate (PPLN) using the CERES, VIBRA epi-, and VIBRA transmission setup. (a)-(c): The raw BCARS spectra of the domain wall (orange) exhibit an additional signal at around \SI{600}{\per\cm} compared to the LNO domain spectrum (blue). (d)-(f): The contrast of the domain walls is imaged by calculating the intensity ratio of the two measured peaks. The out-of-plane arrow symbols indicate the direction of the crystal order parameter. The single spectra (a)-(c) are taken from the maps at the positions indicated by the colored crosses. The PPLN maps were taken with pixel dwell times of \SI{50}{\ms} for CERES, \SI{20}{\ms} for VIBRA epi-, and \SI{1}{\ms} for VIBRA transmission detection. In all cases, the lateral step size is \SI{0.5}{\micro\meter}.
	\label{fig:dw_imaging}}
\end{figure*}

A promising application of BCARS microscopy in crystalline materials is imaging ferroelectric domain walls, as recently shown for LNO \cite{Reitzig2022}. LNO can be poled to locally generate inverted order parameters, distinguishing between original and inverted domains. The boundary region between two domains is called the domain wall (DW), which has many exciting features and applications \cite{Zhao:20, Lu:19}.
SR is an established technique for DW imaging and investigation \cite{Rusing2016a, Rusing2018, Reitzig2021}, and BCARS can offer increased acquisition speed for better applicability in LNO-based device fabrication monitoring.
Here, the reproducibility of BCARS DW imaging is tested in the CERES setup using epi-detection and the VIBRA setup using epi- and transmission-detection. On a single sample of periodically poled LNO, areas of \SI{50}{} $\times$ \SI{50}{\micro\meter\squared} were scanned with a step size of \SI{500}{\nano \meter} in both x- and y-direction. The results are shown in figure \ref{fig:dw_imaging}.

Panels \ref{fig:dw_imaging} (a)-(c) show the raw BCARS spectra of the LNO domain (blue) and the domain wall (orange).
In the domain, the spectra detected in this range exhibit a strong $A_1(LO_4)$ mode at \SI{868}{\per\cm}.
At the DW position, the $A_1(LO_4)$ mode decreases in intensity, while a new mode appears at \SI{600}{\per\cm}, which is assigned to the $A_1(TO_4)$ mode \cite{Stone2012, Rusing2018}.
The ratio between the two peak intensities yields a good contrast for imaging, as seen in panels \ref{fig:dw_imaging} (d)-(f). The yellow-green color indicates the DWs, which delineate the poled domains.
The same vertically aligned cylinder shape can be imaged for all three measurement modes. The scan region was chosen to show the endpoints of the domains and an area of virgin LNO.
The lateral resolution of the DW signal is comparable for all three images. Although the DWs have a thickness of only a few unit cells, the detected signal is broadened to the diffraction-limited spatial resolution.

As the $A_1(TO_4)$ mode substantially varies in intensity, the contrast is visible even for low signal-to-noise ratio (SNR), allowing for fast acquisition speed.
For CERES, the SNR of \SI{13}{} for DW and \SI{17}{} for the domain spectrum limit the pixel dwell time to a minimum of \SI{50}{\ms} while using the maximum laser power of \SI{100}{\mW} for the pump- and \SI{80}{\mW} for the Stokes laser.
In the VIBRA setup, the SNR is higher both in epi- (\SI{29}{}-\SI{69}{}) and transmission direction (\SI{231}{}-\SI{337}{}).
The VIBRA epi measurement was done with \SI{20}{\ms} acquisition time and laser powers of \SI{100}{\mW} and \SI{10}{\mW} for pump- and Stokes-laser, respectively.
The VIBRA transmission measurement has the highest intensity and fastest pixel dwell time of \SI{1}{ms}, which is limited by the detector read-out speed. Additionally, the pump laser power was reduced to \SI{40}{\mW} not to saturate the detector.
Ignoring the detector limitations and using the nonlinear relation between the signal intensity and the laser power of $I_{CARS} \propto I_{pu}^2I{s}$, the pixel dwell time in transmission mode at \SI{100}{\percent} pump power could be reduced to \SI{3.6}{\micro\second} and still reach CERES' levels of SNR. At a pulse rate of \SI{2}{\MHz}, this translates to eight pulses per pixel.
In conclusion, both setups and detection directions show the capability of high-speed imaging of domain walls in LNO. The spectra are reproducible and contain equivalent information. Furthermore, the higher intensity in the forward direction allows for faster imaging, which is limited by detector read-out speed and scan stage movement rather than signal intensity or SNR.

\section{Conclusions and Outlook}

In this work, we have conducted an internal Round Robin by comparing BCARS from two different setups and two different detection modes for analyzing single crystalline materials with increasing complexity: diamond, 6\ch{H-SiC},  KDP, and KTP.
The NRB correction is shown to work equally well for all cases and can remove the setup-dependent influences from the laser source and detector response function. As a result, the phase-retrieved spectra are comparable, reproducible, and in good agreement with SR measurements and literature values.
Both forward- and epi-detected signals contain essentially the same spectral information. The epi-detected signal is shown to be the reflection of the forward signal, as the backward signal is non-detectable due to the phase mismatch between the pump/Stokes and the BCARS signals. The transmission signal is more intense and has a better SNR. Therefore, it is advised to use transmission detection whenever the setup and the sample system allow for it, meaning transparent materials or samples on transmissive substrates. Epi-detection is shown to be equal in information value but can only be used for transparent materials as well, as the reflection of the scattered signal on the backside of the sample is needed.
The comparison of setups showed differences in spectral resolution and signal intensity, depending on the laser system. Furthermore, the imaging of ferroelectric domain walls in LNO was compared in both setups showing the possibility for even faster acquisition speed using transmission detection.

\begin{acknowledgments}The authors gratefully acknowledge financial support by the Deutsche Forschungsgemeinschaft (DFG) through projects \mbox{CRC1415 (ID: 417590517)}, \mbox{INST 269/656-1 FUGG} and FOR5044 (ID: 426703838; \url{http:\\www.For5044.de}), as well as the Würzburg-Dresden Cluster of Excellence” ct.qmat” (EXC 2147), LASERLAB-EUROPE (grant agreement no. 871124, European Union’s Horizon 2020 research and innovation program), and the European Union project CRIMSON (grant agreement no. 101016923).
The authors thank Ji\v{r}\'{i} Hlinka of the Czech Academy of Sciences, Prague, Czech Republic, for providing samples.
\end{acknowledgments}

The authors F. Hempel and F. Vernuccio equally contributed to this work.


\begin{thebibliography}{60}%
\makeatletter
\providecommand \@ifxundefined [1]{%
 \@ifx{#1\undefined}
}%
\providecommand \@ifnum [1]{%
 \ifnum #1\expandafter \@firstoftwo
 \else \expandafter \@secondoftwo
 \fi
}%
\providecommand \@ifx [1]{%
 \ifx #1\expandafter \@firstoftwo
 \else \expandafter \@secondoftwo
 \fi
}%
\providecommand \natexlab [1]{#1}%
\providecommand \enquote  [1]{``#1''}%
\providecommand \bibnamefont  [1]{#1}%
\providecommand \bibfnamefont [1]{#1}%
\providecommand \citenamefont [1]{#1}%
\providecommand \href@noop [0]{\@secondoftwo}%
\providecommand \href [0]{\begingroup \@sanitize@url \@href}%
\providecommand \@href[1]{\@@startlink{#1}\@@href}%
\providecommand \@@href[1]{\endgroup#1\@@endlink}%
\providecommand \@sanitize@url [0]{\catcode `\\12\catcode `\$12\catcode
  `\&12\catcode `\#12\catcode `\^12\catcode `\_12\catcode `\%12\relax}%
\providecommand \@@startlink[1]{}%
\providecommand \@@endlink[0]{}%
\providecommand \url  [0]{\begingroup\@sanitize@url \@url }%
\providecommand \@url [1]{\endgroup\@href {#1}{\urlprefix }}%
\providecommand \urlprefix  [0]{URL }%
\providecommand \Eprint [0]{\href }%
\providecommand \doibase [0]{https://doi.org/}%
\providecommand \selectlanguage [0]{\@gobble}%
\providecommand \bibinfo  [0]{\@secondoftwo}%
\providecommand \bibfield  [0]{\@secondoftwo}%
\providecommand \translation [1]{[#1]}%
\providecommand \BibitemOpen [0]{}%
\providecommand \bibitemStop [0]{}%
\providecommand \bibitemNoStop [0]{.\EOS\space}%
\providecommand \EOS [0]{\spacefactor3000\relax}%
\providecommand \BibitemShut  [1]{\csname bibitem#1\endcsname}%
\let\auto@bib@innerbib\@empty
\bibitem [{\citenamefont {Vanna}\ \emph {et~al.}(2022)\citenamefont {Vanna},
  \citenamefont {{De la Cadena}}, \citenamefont {Talone}, \citenamefont
  {Manzoni}, \citenamefont {Marangoni}, \citenamefont {Polli},\ and\
  \citenamefont {Cerullo}}]{Vanna2022}%
  \BibitemOpen
  \bibfield  {author} {\bibinfo {author} {\bibfnamefont {R.}~\bibnamefont
  {Vanna}}, \bibinfo {author} {\bibfnamefont {A.}~\bibnamefont {{De la
  Cadena}}}, \bibinfo {author} {\bibfnamefont {B.}~\bibnamefont {Talone}},
  \bibinfo {author} {\bibfnamefont {C.}~\bibnamefont {Manzoni}}, \bibinfo
  {author} {\bibfnamefont {M.}~\bibnamefont {Marangoni}}, \bibinfo {author}
  {\bibfnamefont {D.}~\bibnamefont {Polli}},\ and\ \bibinfo {author}
  {\bibfnamefont {G.}~\bibnamefont {Cerullo}},\ }\bibfield  {title} {\bibinfo
  {title} {{Vibrational imaging for label-free cancer diagnosis and
  classification}},\ }\href {https://doi.org/10.1007/s40766-021-00027-6}
  {\bibfield  {journal} {\bibinfo  {journal} {La Rivista del Nuovo Cimento}\
  }\textbf {\bibinfo {volume} {45}},\ \bibinfo {pages} {107} (\bibinfo {year}
  {2022})}\BibitemShut {NoStop}%
\bibitem [{\citenamefont {Turrell}\ and\ \citenamefont
  {Corset}(1996)}]{turrell1996}%
  \BibitemOpen
  \bibfield  {author} {\bibinfo {author} {\bibfnamefont {G.}~\bibnamefont
  {Turrell}}\ and\ \bibinfo {author} {\bibfnamefont {J.}~\bibnamefont
  {Corset}},\ }\href@noop {} {\emph {\bibinfo {title} {{Raman microscopy:
  developments and applications}}}}\ (\bibinfo  {publisher} {Academic Press},\
  \bibinfo {year} {1996})\BibitemShut {NoStop}%
\bibitem [{\citenamefont {Fontana}\ and\ \citenamefont
  {Bourson}(2015)}]{Fontana2015}%
  \BibitemOpen
  \bibfield  {author} {\bibinfo {author} {\bibfnamefont {M.~D.}\ \bibnamefont
  {Fontana}}\ and\ \bibinfo {author} {\bibfnamefont {P.}~\bibnamefont
  {Bourson}},\ }\bibfield  {title} {\bibinfo {title} {{Microstructure and
  defects probed by Raman spectroscopy in lithium niobate crystals and
  devices}},\ }\bibfield  {journal} {\bibinfo  {journal} {Applied Physics
  Reviews}\ }\textbf {\bibinfo {volume} {2}},\ \href
  {https://doi.org/10.1063/1.4934203} {10.1063/1.4934203} (\bibinfo {year}
  {2015})\BibitemShut {NoStop}%
\bibitem [{\citenamefont {Hayazawa}\ \emph {et~al.}(2007)\citenamefont
  {Hayazawa}, \citenamefont {Motohashi}, \citenamefont {Saito}, \citenamefont
  {Ishitobi}, \citenamefont {Ono}, \citenamefont {Ichimura}, \citenamefont
  {Verma},\ and\ \citenamefont {Kawata}}]{hayazawa2007}%
  \BibitemOpen
  \bibfield  {author} {\bibinfo {author} {\bibfnamefont {N.}~\bibnamefont
  {Hayazawa}}, \bibinfo {author} {\bibfnamefont {M.}~\bibnamefont {Motohashi}},
  \bibinfo {author} {\bibfnamefont {Y.}~\bibnamefont {Saito}}, \bibinfo
  {author} {\bibfnamefont {H.}~\bibnamefont {Ishitobi}}, \bibinfo {author}
  {\bibfnamefont {A.}~\bibnamefont {Ono}}, \bibinfo {author} {\bibfnamefont
  {T.}~\bibnamefont {Ichimura}}, \bibinfo {author} {\bibfnamefont
  {P.}~\bibnamefont {Verma}},\ and\ \bibinfo {author} {\bibfnamefont
  {S.}~\bibnamefont {Kawata}},\ }\bibfield  {title} {\bibinfo {title}
  {{Visualization of localized strain of a crystalline thin layer at the
  nanoscale by tip-enhanced Raman spectroscopy and microscopy}},\ }\href@noop
  {} {\bibfield  {journal} {\bibinfo  {journal} {Journal of Raman Spectroscopy:
  An International Journal for Original Work in all Aspects of Raman
  Spectroscopy, Including Higher Order Processes, and also Brillouin and
  Rayleigh Scattering}\ }\textbf {\bibinfo {volume} {38}},\ \bibinfo {pages}
  {684} (\bibinfo {year} {2007})}\BibitemShut {NoStop}%
\bibitem [{\citenamefont {Prawer}\ and\ \citenamefont
  {Nemanich}(2004)}]{prawer2004}%
  \BibitemOpen
  \bibfield  {author} {\bibinfo {author} {\bibfnamefont {S.}~\bibnamefont
  {Prawer}}\ and\ \bibinfo {author} {\bibfnamefont {R.~J.}\ \bibnamefont
  {Nemanich}},\ }\bibfield  {title} {\bibinfo {title} {{Raman spectroscopy of
  diamond and doped diamond}},\ }\href@noop {} {\bibfield  {journal} {\bibinfo
  {journal} {Philosophical Transactions of the Royal Society of London. Series
  A: Mathematical, Physical and Engineering Sciences}\ }\textbf {\bibinfo
  {volume} {362}},\ \bibinfo {pages} {2537} (\bibinfo {year}
  {2004})}\BibitemShut {NoStop}%
\bibitem [{\citenamefont {Polli}\ \emph {et~al.}(2018)\citenamefont {Polli},
  \citenamefont {Kumar}, \citenamefont {Valensise}, \citenamefont {Marangoni},\
  and\ \citenamefont {Cerullo}}]{Polli2018}%
  \BibitemOpen
  \bibfield  {author} {\bibinfo {author} {\bibfnamefont {D.}~\bibnamefont
  {Polli}}, \bibinfo {author} {\bibfnamefont {V.}~\bibnamefont {Kumar}},
  \bibinfo {author} {\bibfnamefont {C.~M.}\ \bibnamefont {Valensise}}, \bibinfo
  {author} {\bibfnamefont {M.}~\bibnamefont {Marangoni}},\ and\ \bibinfo
  {author} {\bibfnamefont {G.}~\bibnamefont {Cerullo}},\ }\bibfield  {title}
  {\bibinfo {title} {{Broadband Coherent Raman Scattering Microscopy}},\ }\href
  {https://doi.org/10.1002/lpor.201800020} {\bibfield  {journal} {\bibinfo
  {journal} {Laser and Photonics Reviews}\ }\textbf {\bibinfo {volume} {12}},\
  \bibinfo {pages} {1800020} (\bibinfo {year} {2018})}\BibitemShut {NoStop}%
\bibitem [{\citenamefont {Cheng}\ and\ \citenamefont {Xie}(2016)}]{cheng2016}%
  \BibitemOpen
  \bibfield  {author} {\bibinfo {author} {\bibfnamefont {J.-X.}\ \bibnamefont
  {Cheng}}\ and\ \bibinfo {author} {\bibfnamefont {X.~S.}\ \bibnamefont
  {Xie}},\ }\href@noop {} {\emph {\bibinfo {title} {{Coherent Raman scattering
  microscopy}}}}\ (\bibinfo  {publisher} {CRC press},\ \bibinfo {year}
  {2016})\BibitemShut {NoStop}%
\bibitem [{\citenamefont {Kolesnichenko}\ \emph {et~al.}(2019)\citenamefont
  {Kolesnichenko}, \citenamefont {Tollerud},\ and\ \citenamefont
  {Davis}}]{Kolesnichenko2019}%
  \BibitemOpen
  \bibfield  {author} {\bibinfo {author} {\bibfnamefont {P.~V.}\ \bibnamefont
  {Kolesnichenko}}, \bibinfo {author} {\bibfnamefont {J.~O.}\ \bibnamefont
  {Tollerud}},\ and\ \bibinfo {author} {\bibfnamefont {J.~A.}\ \bibnamefont
  {Davis}},\ }\bibfield  {title} {\bibinfo {title} {{Background-free
  time-resolved coherent Raman spectroscopy (CSRS and CARS): Heterodyne
  detection of low-energy vibrations and identification of excited-state
  contributions}},\ }\bibfield  {journal} {\bibinfo  {journal} {APL Photonics}\
  }\textbf {\bibinfo {volume} {4}},\ \href {https://doi.org/10.1063/1.5090585}
  {10.1063/1.5090585} (\bibinfo {year} {2019})\BibitemShut {NoStop}%
\bibitem [{\citenamefont {Cheng}\ \emph {et~al.}(2001)\citenamefont {Cheng},
  \citenamefont {Book},\ and\ \citenamefont {Xie}}]{Cheng:01}%
  \BibitemOpen
  \bibfield  {author} {\bibinfo {author} {\bibfnamefont {J.-X.}\ \bibnamefont
  {Cheng}}, \bibinfo {author} {\bibfnamefont {L.~D.}\ \bibnamefont {Book}},\
  and\ \bibinfo {author} {\bibfnamefont {X.~S.}\ \bibnamefont {Xie}},\
  }\bibfield  {title} {\bibinfo {title} {{Polarization coherent anti-Stokes
  Raman scattering microscopy}},\ }\href {https://doi.org/10.1364/OL.26.001341}
  {\bibfield  {journal} {\bibinfo  {journal} {Opt. Lett.}\ }\textbf {\bibinfo
  {volume} {26}},\ \bibinfo {pages} {1341} (\bibinfo {year}
  {2001})}\BibitemShut {NoStop}%
\bibitem [{\citenamefont {Cui}\ \emph {et~al.}(2006)\citenamefont {Cui},
  \citenamefont {Joffre}, \citenamefont {Skodack},\ and\ \citenamefont
  {Ogilvie}}]{Cui:06}%
  \BibitemOpen
  \bibfield  {author} {\bibinfo {author} {\bibfnamefont {M.}~\bibnamefont
  {Cui}}, \bibinfo {author} {\bibfnamefont {M.}~\bibnamefont {Joffre}},
  \bibinfo {author} {\bibfnamefont {J.}~\bibnamefont {Skodack}},\ and\ \bibinfo
  {author} {\bibfnamefont {J.~P.}\ \bibnamefont {Ogilvie}},\ }\bibfield
  {title} {\bibinfo {title} {{Interferometric Fourier transform coherent
  anti-stokes Raman scattering}},\ }\href
  {https://doi.org/10.1364/OE.14.008448} {\bibfield  {journal} {\bibinfo
  {journal} {Opt. Express}\ }\textbf {\bibinfo {volume} {14}},\ \bibinfo
  {pages} {8448} (\bibinfo {year} {2006})}\BibitemShut {NoStop}%
\bibitem [{\citenamefont {Cheng}\ \emph {et~al.}(2002)\citenamefont {Cheng},
  \citenamefont {Volkmer},\ and\ \citenamefont {Xie}}]{cheng2002}%
  \BibitemOpen
  \bibfield  {author} {\bibinfo {author} {\bibfnamefont {J.-X.}\ \bibnamefont
  {Cheng}}, \bibinfo {author} {\bibfnamefont {A.}~\bibnamefont {Volkmer}},\
  and\ \bibinfo {author} {\bibfnamefont {X.~S.}\ \bibnamefont {Xie}},\
  }\bibfield  {title} {\bibinfo {title} {{Theoretical and experimental
  characterization of coherent anti-Stokes Raman scattering microscopy}},\
  }\href {https://doi.org/10.1364/josab.19.001363} {\bibfield  {journal}
  {\bibinfo  {journal} {Journal of the Optical Society of America B}\ }\textbf
  {\bibinfo {volume} {19}},\ \bibinfo {pages} {1363} (\bibinfo {year}
  {2002})}\BibitemShut {NoStop}%
\bibitem [{\citenamefont {Vartiainen}(1992)}]{Vartiainen1992}%
  \BibitemOpen
  \bibfield  {author} {\bibinfo {author} {\bibfnamefont {E.~M.}\ \bibnamefont
  {Vartiainen}},\ }\bibfield  {title} {\bibinfo {title} {{Phase retrieval
  approach for coherent anti-Stokes Raman scattering spectrum analysis}},\
  }\href {https://doi.org/10.1364/JOSAB.9.001209} {\bibfield  {journal}
  {\bibinfo  {journal} {Journal of the Optical Society of America B}\ }\textbf
  {\bibinfo {volume} {9}},\ \bibinfo {pages} {1209} (\bibinfo {year}
  {1992})}\BibitemShut {NoStop}%
\bibitem [{\citenamefont {Camp}\ \emph {et~al.}(2016)\citenamefont {Camp},
  \citenamefont {Lee},\ and\ \citenamefont {Cicerone}}]{Camp2016}%
  \BibitemOpen
  \bibfield  {author} {\bibinfo {author} {\bibfnamefont {C.~H.}\ \bibnamefont
  {Camp}}, \bibinfo {author} {\bibfnamefont {Y.~J.}\ \bibnamefont {Lee}},\ and\
  \bibinfo {author} {\bibfnamefont {M.~T.}\ \bibnamefont {Cicerone}},\
  }\bibfield  {title} {\bibinfo {title} {{Quantitative, comparable coherent
  anti-Stokes Raman scattering (CARS) spectroscopy: Correcting errors in phase
  retrieval}},\ }\href {https://doi.org/10.1002/jrs.4824} {\bibfield  {journal}
  {\bibinfo  {journal} {Journal of Raman Spectroscopy}\ }\textbf {\bibinfo
  {volume} {47}},\ \bibinfo {pages} {408} (\bibinfo {year} {2016})},\ \Eprint
  {https://arxiv.org/abs/arXiv:1507.06543v1} {arXiv:arXiv:1507.06543v1}
  \BibitemShut {NoStop}%
\bibitem [{\citenamefont {Valensise}\ \emph {et~al.}(2020)\citenamefont
  {Valensise}, \citenamefont {Giuseppi}, \citenamefont {Vernuccio},
  \citenamefont {la~Cadena}, \citenamefont {Cerullo},\ and\ \citenamefont
  {Polli}}]{Valensise2020}%
  \BibitemOpen
  \bibfield  {author} {\bibinfo {author} {\bibfnamefont {C.~M.}\ \bibnamefont
  {Valensise}}, \bibinfo {author} {\bibfnamefont {A.}~\bibnamefont {Giuseppi}},
  \bibinfo {author} {\bibfnamefont {F.}~\bibnamefont {Vernuccio}}, \bibinfo
  {author} {\bibfnamefont {A.}~\bibnamefont {la~Cadena}}, \bibinfo {author}
  {\bibfnamefont {G.}~\bibnamefont {Cerullo}},\ and\ \bibinfo {author}
  {\bibfnamefont {D.}~\bibnamefont {Polli}},\ }\bibfield  {title} {\bibinfo
  {title} {{Removing non-resonant background from CARS spectra via deep
  learning}},\ }\href {https://doi.org/10.1063/5.0007821} {\bibfield  {journal}
  {\bibinfo  {journal} {APL Photonics}\ }\textbf {\bibinfo {volume} {5}},\
  \bibinfo {pages} {61305} (\bibinfo {year} {2020})}\BibitemShut {NoStop}%
\bibitem [{\citenamefont {Houhou}\ \emph {et~al.}(2020)\citenamefont {Houhou},
  \citenamefont {Barman}, \citenamefont {Schmitt}, \citenamefont {Meyer},
  \citenamefont {Popp},\ and\ \citenamefont {Bocklitz}}]{Houhou2020}%
  \BibitemOpen
  \bibfield  {author} {\bibinfo {author} {\bibfnamefont {R.}~\bibnamefont
  {Houhou}}, \bibinfo {author} {\bibfnamefont {P.}~\bibnamefont {Barman}},
  \bibinfo {author} {\bibfnamefont {M.}~\bibnamefont {Schmitt}}, \bibinfo
  {author} {\bibfnamefont {T.}~\bibnamefont {Meyer}}, \bibinfo {author}
  {\bibfnamefont {J.}~\bibnamefont {Popp}},\ and\ \bibinfo {author}
  {\bibfnamefont {T.}~\bibnamefont {Bocklitz}},\ }\bibfield  {title}
  {{\bibinfo {title} {{Deep learning as phase retrieval
  tool for CARS spectra}}},\ }\href
  {https://doi.org/https://doi.org/10.1364/OE.390413} {\bibfield  {journal}
  {\bibinfo  {journal} {Optics express : the international electronic journal
  of optics}\ }\textbf {\bibinfo {volume} {28}},\ \bibinfo {pages} {21002}
  (\bibinfo {year} {2020})}\BibitemShut {NoStop}%
\bibitem [{\citenamefont {Wang}\ \emph {et~al.}(2022)\citenamefont {Wang},
  \citenamefont {{O' Dwyer}}, \citenamefont {Muddiman}, \citenamefont {Ward},
  \citenamefont {{Camp Jr.}},\ and\ \citenamefont {Hennelly}}]{Wang2022}%
  \BibitemOpen
  \bibfield  {author} {\bibinfo {author} {\bibfnamefont {Z.}~\bibnamefont
  {Wang}}, \bibinfo {author} {\bibfnamefont {K.}~\bibnamefont {{O' Dwyer}}},
  \bibinfo {author} {\bibfnamefont {R.}~\bibnamefont {Muddiman}}, \bibinfo
  {author} {\bibfnamefont {T.}~\bibnamefont {Ward}}, \bibinfo {author}
  {\bibfnamefont {C.~H.}\ \bibnamefont {{Camp Jr.}}},\ and\ \bibinfo {author}
  {\bibfnamefont {B.~M.}\ \bibnamefont {Hennelly}},\ }\bibfield  {title}
  {\bibinfo {title} {{VECTOR: Very deep convolutional autoencoders for
  non-resonant background removal in broadband coherent anti-Stokes Raman
  scattering}},\ }\href {https://doi.org/https://doi.org/10.1002/jrs.6335}
  {\bibfield  {journal} {\bibinfo  {journal} {Journal of Raman Spectroscopy}\
  }\textbf {\bibinfo {volume} {53}},\ \bibinfo {pages} {1081} (\bibinfo {year}
  {2022})}\BibitemShut {NoStop}%
\bibitem [{\citenamefont {{De La Cadena}}\ \emph {et~al.}(2022)\citenamefont
  {{De La Cadena}}, \citenamefont {Vernuccio}, \citenamefont {Ragni},
  \citenamefont {Sciortino}, \citenamefont {Vanna}, \citenamefont {Ferrante},
  \citenamefont {Pediconi}, \citenamefont {Valensise}, \citenamefont {Genchi},
  \citenamefont {Laptenok}, \citenamefont {Doni}, \citenamefont {Erreni},
  \citenamefont {Scopigno}, \citenamefont {Liberale}, \citenamefont {Ferrari},
  \citenamefont {Sampietro}, \citenamefont {Cerullo},\ and\ \citenamefont
  {Polli}}]{Delacadena2022}%
  \BibitemOpen
  \bibfield  {author} {\bibinfo {author} {\bibfnamefont {A.}~\bibnamefont {{De
  La Cadena}}}, \bibinfo {author} {\bibfnamefont {F.}~\bibnamefont
  {Vernuccio}}, \bibinfo {author} {\bibfnamefont {A.}~\bibnamefont {Ragni}},
  \bibinfo {author} {\bibfnamefont {G.}~\bibnamefont {Sciortino}}, \bibinfo
  {author} {\bibfnamefont {R.}~\bibnamefont {Vanna}}, \bibinfo {author}
  {\bibfnamefont {C.}~\bibnamefont {Ferrante}}, \bibinfo {author}
  {\bibfnamefont {N.}~\bibnamefont {Pediconi}}, \bibinfo {author}
  {\bibfnamefont {C.}~\bibnamefont {Valensise}}, \bibinfo {author}
  {\bibfnamefont {L.}~\bibnamefont {Genchi}}, \bibinfo {author} {\bibfnamefont
  {S.~P.}\ \bibnamefont {Laptenok}}, \bibinfo {author} {\bibfnamefont
  {A.}~\bibnamefont {Doni}}, \bibinfo {author} {\bibfnamefont {M.}~\bibnamefont
  {Erreni}}, \bibinfo {author} {\bibfnamefont {T.}~\bibnamefont {Scopigno}},
  \bibinfo {author} {\bibfnamefont {C.}~\bibnamefont {Liberale}}, \bibinfo
  {author} {\bibfnamefont {G.}~\bibnamefont {Ferrari}}, \bibinfo {author}
  {\bibfnamefont {M.}~\bibnamefont {Sampietro}}, \bibinfo {author}
  {\bibfnamefont {G.}~\bibnamefont {Cerullo}},\ and\ \bibinfo {author}
  {\bibfnamefont {D.}~\bibnamefont {Polli}},\ }\bibfield  {title} {\bibinfo
  {title} {{Broadband stimulated Raman imaging based on multi-channel lock-in
  detection for spectral histopathology}},\ }\bibfield  {journal} {\bibinfo
  {journal} {APL Photonics}\ }\textbf {\bibinfo {volume} {7}},\ \href
  {https://doi.org/10.1063/5.0093946} {10.1063/5.0093946} (\bibinfo {year}
  {2022})\BibitemShut {NoStop}%
\bibitem [{\citenamefont {Cicerone}(2016)}]{Cicerone2016}%
  \BibitemOpen
  \bibfield  {author} {\bibinfo {author} {\bibfnamefont {M.}~\bibnamefont
  {Cicerone}},\ }\bibfield  {title} {\bibinfo {title} {{Molecular imaging with
  CARS micro-spectroscopy}},\ }\href
  {https://doi.org/10.1016/j.cbpa.2016.05.010} {\bibfield  {journal} {\bibinfo
  {journal} {Current Opinion in Chemical Biology}\ }\textbf {\bibinfo {volume}
  {33}},\ \bibinfo {pages} {179} (\bibinfo {year} {2016})}\BibitemShut
  {NoStop}%
\bibitem [{\citenamefont {Vernuccio}\ \emph {et~al.}(2023)\citenamefont
  {Vernuccio}, \citenamefont {Vanna}, \citenamefont {Ceconello}, \citenamefont
  {Bresci}, \citenamefont {Manetti}, \citenamefont {Sorrentino}, \citenamefont
  {Ghislanzoni}, \citenamefont {Lambertucci}, \citenamefont {Moti{\~{n}}o},
  \citenamefont {Martins}, \citenamefont {Kroemer}, \citenamefont {Bongarzone},
  \citenamefont {Cerullo},\ and\ \citenamefont {Polli}}]{Vernuccio2023}%
  \BibitemOpen
  \bibfield  {author} {\bibinfo {author} {\bibfnamefont {F.}~\bibnamefont
  {Vernuccio}}, \bibinfo {author} {\bibfnamefont {R.}~\bibnamefont {Vanna}},
  \bibinfo {author} {\bibfnamefont {C.}~\bibnamefont {Ceconello}}, \bibinfo
  {author} {\bibfnamefont {A.}~\bibnamefont {Bresci}}, \bibinfo {author}
  {\bibfnamefont {F.}~\bibnamefont {Manetti}}, \bibinfo {author} {\bibfnamefont
  {S.}~\bibnamefont {Sorrentino}}, \bibinfo {author} {\bibfnamefont
  {S.}~\bibnamefont {Ghislanzoni}}, \bibinfo {author} {\bibfnamefont
  {F.}~\bibnamefont {Lambertucci}}, \bibinfo {author} {\bibfnamefont
  {O.}~\bibnamefont {Moti{\~{n}}o}}, \bibinfo {author} {\bibfnamefont
  {I.}~\bibnamefont {Martins}}, \bibinfo {author} {\bibfnamefont
  {G.}~\bibnamefont {Kroemer}}, \bibinfo {author} {\bibfnamefont
  {I.}~\bibnamefont {Bongarzone}}, \bibinfo {author} {\bibfnamefont
  {G.}~\bibnamefont {Cerullo}},\ and\ \bibinfo {author} {\bibfnamefont
  {D.}~\bibnamefont {Polli}},\ }\bibfield  {title} {\bibinfo {title}
  {{Full-spectrum CARS Microscopy Of Cells And Tissues With Ultrashort
  White-light Continuum Pulses}},\ }\bibfield  {journal} {\bibinfo  {journal}
  {J. Phys. Chem. B}\ }\href {https://doi.org/10.1021/acs.jpcb.3c01443}
  {10.1021/acs.jpcb.3c01443} (\bibinfo {year} {2023})\BibitemShut {NoStop}%
\bibitem [{\citenamefont {Pope}\ \emph {et~al.}(2014)\citenamefont {Pope},
  \citenamefont {Payne}, \citenamefont {Zoriniants}, \citenamefont {Thomas},
  \citenamefont {Williams}, \citenamefont {Watson}, \citenamefont {Langbein},\
  and\ \citenamefont {Borri}}]{Pope2014}%
  \BibitemOpen
  \bibfield  {author} {\bibinfo {author} {\bibfnamefont {I.}~\bibnamefont
  {Pope}}, \bibinfo {author} {\bibfnamefont {L.}~\bibnamefont {Payne}},
  \bibinfo {author} {\bibfnamefont {G.}~\bibnamefont {Zoriniants}}, \bibinfo
  {author} {\bibfnamefont {E.}~\bibnamefont {Thomas}}, \bibinfo {author}
  {\bibfnamefont {O.}~\bibnamefont {Williams}}, \bibinfo {author}
  {\bibfnamefont {P.}~\bibnamefont {Watson}}, \bibinfo {author} {\bibfnamefont
  {W.}~\bibnamefont {Langbein}},\ and\ \bibinfo {author} {\bibfnamefont
  {P.}~\bibnamefont {Borri}},\ }\bibfield  {title} {\bibinfo {title} {{Coherent
  anti-Stokes Raman scattering microscopy of single nanodiamonds}},\ }\href
  {https://doi.org/10.1038/nnano.2014.210} {\bibfield  {journal} {\bibinfo
  {journal} {Nature Nanotechnology}\ }\textbf {\bibinfo {volume} {9}},\
  \bibinfo {pages} {940} (\bibinfo {year} {2014})}\BibitemShut {NoStop}%
\bibitem [{\citenamefont {Hempel}\ \emph {et~al.}(2021)\citenamefont {Hempel},
  \citenamefont {Reitzig}, \citenamefont {R{\"{u}}sing},\ and\ \citenamefont
  {Eng}}]{Hempel2021}%
  \BibitemOpen
  \bibfield  {author} {\bibinfo {author} {\bibfnamefont {F.}~\bibnamefont
  {Hempel}}, \bibinfo {author} {\bibfnamefont {S.}~\bibnamefont {Reitzig}},
  \bibinfo {author} {\bibfnamefont {M.}~\bibnamefont {R{\"{u}}sing}},\ and\
  \bibinfo {author} {\bibfnamefont {L.~M.}\ \bibnamefont {Eng}},\ }\bibfield
  {title} {\bibinfo {title} {{Broadband coherent anti-Stokes Raman scattering
  for crystalline materials}},\ }\href
  {https://doi.org/10.1103/PhysRevB.104.224308} {\bibfield  {journal} {\bibinfo
   {journal} {Physical Review B}\ }\textbf {\bibinfo {volume} {104}},\ \bibinfo
  {pages} {224308} (\bibinfo {year} {2021})}\BibitemShut {NoStop}%
\bibitem [{\citenamefont {Reitzig}\ \emph {et~al.}(2022)\citenamefont
  {Reitzig}, \citenamefont {Hempel}, \citenamefont {Ratzenberger},
  \citenamefont {Hegarty}, \citenamefont {Amber}, \citenamefont {Buschbeck},
  \citenamefont {R{\"{u}}sing},\ and\ \citenamefont {Eng}}]{Reitzig2022}%
  \BibitemOpen
  \bibfield  {author} {\bibinfo {author} {\bibfnamefont {S.}~\bibnamefont
  {Reitzig}}, \bibinfo {author} {\bibfnamefont {F.}~\bibnamefont {Hempel}},
  \bibinfo {author} {\bibfnamefont {J.}~\bibnamefont {Ratzenberger}}, \bibinfo
  {author} {\bibfnamefont {P.~A.}\ \bibnamefont {Hegarty}}, \bibinfo {author}
  {\bibfnamefont {Z.~H.}\ \bibnamefont {Amber}}, \bibinfo {author}
  {\bibfnamefont {R.}~\bibnamefont {Buschbeck}}, \bibinfo {author}
  {\bibfnamefont {M.}~\bibnamefont {R{\"{u}}sing}},\ and\ \bibinfo {author}
  {\bibfnamefont {L.~M.}\ \bibnamefont {Eng}},\ }\bibfield  {title} {\bibinfo
  {title} {{High-speed hyperspectral imaging of ferroelectric domain walls
  using broadband coherent anti-Stokes Raman scattering}},\ }\href
  {https://doi.org/10.1063/5.0086029} {\bibfield  {journal} {\bibinfo
  {journal} {Applied Physics Letters}\ }\textbf {\bibinfo {volume} {120}},\
  \bibinfo {pages} {162901} (\bibinfo {year} {2022})}\BibitemShut {NoStop}%
\bibitem [{\citenamefont {Evans}\ and\ \citenamefont {Xie}(2008)}]{evans2008}%
  \BibitemOpen
  \bibfield  {author} {\bibinfo {author} {\bibfnamefont {C.~L.}\ \bibnamefont
  {Evans}}\ and\ \bibinfo {author} {\bibfnamefont {X.~S.}\ \bibnamefont
  {Xie}},\ }\bibfield  {title} {\bibinfo {title} {{Coherent Anti-Stokes Raman
  Scattering Microscopy: Chemical Imaging for Biology and Medicine}},\ }\href
  {https://doi.org/10.1146/annurev.anchem.1.031207.112754} {\bibfield
  {journal} {\bibinfo  {journal} {Annual Review of Analytical Chemistry}\
  }\textbf {\bibinfo {volume} {1}},\ \bibinfo {pages} {883} (\bibinfo {year}
  {2008})}\BibitemShut {NoStop}%
\bibitem [{\citenamefont {Petrov}\ \emph {et~al.}(2021)\citenamefont {Petrov},
  \citenamefont {Arora},\ and\ \citenamefont {Yakovlev}}]{petrov2021}%
  \BibitemOpen
  \bibfield  {author} {\bibinfo {author} {\bibfnamefont {G.~I.}\ \bibnamefont
  {Petrov}}, \bibinfo {author} {\bibfnamefont {R.}~\bibnamefont {Arora}},\ and\
  \bibinfo {author} {\bibfnamefont {V.~V.}\ \bibnamefont {Yakovlev}},\
  }\bibfield  {title} {\bibinfo {title} {{Coherent anti-Stokes Raman scattering
  imaging of microcalcifications associated with breast cancer}},\ }\href
  {https://doi.org/10.1039/d0an01962c} {\bibfield  {journal} {\bibinfo
  {journal} {Analyst}\ }\textbf {\bibinfo {volume} {146}},\ \bibinfo {pages}
  {1253} (\bibinfo {year} {2021})}\BibitemShut {NoStop}%
\bibitem [{\citenamefont {Zhang}\ \emph {et~al.}(2015)\citenamefont {Zhang},
  \citenamefont {Zhang},\ and\ \citenamefont {Cheng}}]{Zhang2015}%
  \BibitemOpen
  \bibfield  {author} {\bibinfo {author} {\bibfnamefont {C.}~\bibnamefont
  {Zhang}}, \bibinfo {author} {\bibfnamefont {D.}~\bibnamefont {Zhang}},\ and\
  \bibinfo {author} {\bibfnamefont {J.-X.~X.}\ \bibnamefont {Cheng}},\
  }\bibfield  {title} {\bibinfo {title} {{Coherent Raman Scattering Microscopy
  in Biology and Medicine}},\ }\href
  {https://doi.org/10.1146/annurev-bioeng-071114-040554} {\bibfield  {journal}
  {\bibinfo  {journal} {Annual Review of Biomedical Engineering}\ }\textbf
  {\bibinfo {volume} {17}},\ \bibinfo {pages} {415} (\bibinfo {year}
  {2015})}\BibitemShut {NoStop}%
\bibitem [{\citenamefont {Late}\ \emph {et~al.}(2011)\citenamefont {Late},
  \citenamefont {Maitra}, \citenamefont {Panchakarla}, \citenamefont
  {Waghmare},\ and\ \citenamefont {Rao}}]{late2011}%
  \BibitemOpen
  \bibfield  {author} {\bibinfo {author} {\bibfnamefont {D.~J.}\ \bibnamefont
  {Late}}, \bibinfo {author} {\bibfnamefont {U.}~\bibnamefont {Maitra}},
  \bibinfo {author} {\bibfnamefont {L.~S.}\ \bibnamefont {Panchakarla}},
  \bibinfo {author} {\bibfnamefont {U.~V.}\ \bibnamefont {Waghmare}},\ and\
  \bibinfo {author} {\bibfnamefont {C.~N.~R.}\ \bibnamefont {Rao}},\ }\bibfield
   {title} {\bibinfo {title} {{Temperature effects on the Raman spectra of
  graphenes: dependence on the number of layers and doping}},\ }\href@noop {}
  {\bibfield  {journal} {\bibinfo  {journal} {Journal of Physics: Condensed
  Matter}\ }\textbf {\bibinfo {volume} {23}},\ \bibinfo {pages} {55303}
  (\bibinfo {year} {2011})}\BibitemShut {NoStop}%
\bibitem [{\citenamefont {Ferrari}\ \emph {et~al.}(2006)\citenamefont
  {Ferrari}, \citenamefont {Meyer}, \citenamefont {Scardaci}, \citenamefont
  {Casiraghi}, \citenamefont {Lazzeri}, \citenamefont {Mauri}, \citenamefont
  {Piscanec}, \citenamefont {Jiang}, \citenamefont {Novoselov}, \citenamefont
  {Roth},\ and\ \citenamefont {Geim}}]{ferrari2006}%
  \BibitemOpen
  \bibfield  {author} {\bibinfo {author} {\bibfnamefont {A.~C.}\ \bibnamefont
  {Ferrari}}, \bibinfo {author} {\bibfnamefont {J.~C.}\ \bibnamefont {Meyer}},
  \bibinfo {author} {\bibfnamefont {V.}~\bibnamefont {Scardaci}}, \bibinfo
  {author} {\bibfnamefont {C.}~\bibnamefont {Casiraghi}}, \bibinfo {author}
  {\bibfnamefont {M.}~\bibnamefont {Lazzeri}}, \bibinfo {author} {\bibfnamefont
  {F.}~\bibnamefont {Mauri}}, \bibinfo {author} {\bibfnamefont
  {S.}~\bibnamefont {Piscanec}}, \bibinfo {author} {\bibfnamefont
  {D.}~\bibnamefont {Jiang}}, \bibinfo {author} {\bibfnamefont {K.~S.}\
  \bibnamefont {Novoselov}}, \bibinfo {author} {\bibfnamefont {S.}~\bibnamefont
  {Roth}},\ and\ \bibinfo {author} {\bibfnamefont {A.~K.}\ \bibnamefont
  {Geim}},\ }\bibfield  {title} {\bibinfo {title} {{Raman spectrum of graphene
  and graphene layers}},\ }\href
  {https://doi.org/10.1103/PhysRevLett.97.187401} {\bibfield  {journal}
  {\bibinfo  {journal} {Physical Review Letters}\ }\textbf {\bibinfo {volume}
  {97}},\ \bibinfo {pages} {1} (\bibinfo {year} {2006})}\BibitemShut {NoStop}%
\bibitem [{\citenamefont {Guo}\ \emph {et~al.}(2020)\citenamefont {Guo},
  \citenamefont {Beleites}, \citenamefont {Neugebauer}, \citenamefont
  {Abalde-Cela}, \citenamefont {Afseth}, \citenamefont {Alsamad}, \citenamefont
  {Anand}, \citenamefont {Araujo-Andrade}, \citenamefont
  {A{\v{s}}krabi{\'{c}}}, \citenamefont {Avci}, \citenamefont {Baia},
  \citenamefont {Baranska}, \citenamefont {Baria}, \citenamefont {{Batista de
  Carvalho}}, \citenamefont {de~Bettignies}, \citenamefont {Bonifacio},
  \citenamefont {Bonnier}, \citenamefont {Brauchle}, \citenamefont {Byrne},
  \citenamefont {Chourpa}, \citenamefont {Cicchi}, \citenamefont {Cuisinier},
  \citenamefont {Culha}, \citenamefont {Dahms}, \citenamefont {David},
  \citenamefont {Duponchel}, \citenamefont {Duraipandian}, \citenamefont
  {El-Mashtoly}, \citenamefont {Ellis}, \citenamefont {Eppe}, \citenamefont
  {Falgayrac}, \citenamefont {Gamulin}, \citenamefont {Gardner}, \citenamefont
  {Gardner}, \citenamefont {Gerwert}, \citenamefont {Giamarellos-Bourboulis},
  \citenamefont {Gizurarson}, \citenamefont {Gnyba}, \citenamefont {Goodacre},
  \citenamefont {Grysan}, \citenamefont {Guntinas-Lichius}, \citenamefont
  {Helgadottir}, \citenamefont {Gro{\v{s}}ev}, \citenamefont {Kendall},
  \citenamefont {Kiselev}, \citenamefont {K{\"{o}}lbach}, \citenamefont
  {Krafft}, \citenamefont {Krishnamoorthy}, \citenamefont {Kubryck},
  \citenamefont {Lendl}, \citenamefont {Loza-Alvarez}, \citenamefont {Lyng},
  \citenamefont {Machill}, \citenamefont {Malherbe}, \citenamefont {Marro},
  \citenamefont {Marques}, \citenamefont {Matuszyk}, \citenamefont {Morasso},
  \citenamefont {Moreau}, \citenamefont {Muhamadali}, \citenamefont {Mussi},
  \citenamefont {Notingher}, \citenamefont {Pacia}, \citenamefont {Pavone},
  \citenamefont {Penel}, \citenamefont {Petersen}, \citenamefont {Piot},
  \citenamefont {Rau}, \citenamefont {Richter}, \citenamefont {Rybarczyk},
  \citenamefont {Salehi}, \citenamefont {Schenke-Layland}, \citenamefont
  {Schl{\"{u}}cker}, \citenamefont {Schosserer}, \citenamefont {Sch{\"{u}}tze},
  \citenamefont {Sergo}, \citenamefont {Sinjab}, \citenamefont {Smulko},
  \citenamefont {Sockalingum}, \citenamefont {Stiebing}, \citenamefont {Stone},
  \citenamefont {Untereiner}, \citenamefont {Vanna}, \citenamefont {Wieland},
  \citenamefont {Popp},\ and\ \citenamefont {Bocklitz}}]{Guo2020}%
  \BibitemOpen
  \bibfield  {author} {\bibinfo {author} {\bibfnamefont {S.}~\bibnamefont
  {Guo}}, \bibinfo {author} {\bibfnamefont {C.}~\bibnamefont {Beleites}},
  \bibinfo {author} {\bibfnamefont {U.}~\bibnamefont {Neugebauer}}, \bibinfo
  {author} {\bibfnamefont {S.}~\bibnamefont {Abalde-Cela}}, \bibinfo {author}
  {\bibfnamefont {N.~K.}\ \bibnamefont {Afseth}}, \bibinfo {author}
  {\bibfnamefont {F.}~\bibnamefont {Alsamad}}, \bibinfo {author} {\bibfnamefont
  {S.}~\bibnamefont {Anand}}, \bibinfo {author} {\bibfnamefont
  {C.}~\bibnamefont {Araujo-Andrade}}, \bibinfo {author} {\bibfnamefont
  {S.}~\bibnamefont {A{\v{s}}krabi{\'{c}}}}, \bibinfo {author} {\bibfnamefont
  {E.}~\bibnamefont {Avci}}, \bibinfo {author} {\bibfnamefont {M.}~\bibnamefont
  {Baia}}, \bibinfo {author} {\bibfnamefont {M.}~\bibnamefont {Baranska}},
  \bibinfo {author} {\bibfnamefont {E.}~\bibnamefont {Baria}}, \bibinfo
  {author} {\bibfnamefont {L.~A.~E.}\ \bibnamefont {{Batista de Carvalho}}},
  \bibinfo {author} {\bibfnamefont {P.}~\bibnamefont {de~Bettignies}}, \bibinfo
  {author} {\bibfnamefont {A.}~\bibnamefont {Bonifacio}}, \bibinfo {author}
  {\bibfnamefont {F.}~\bibnamefont {Bonnier}}, \bibinfo {author} {\bibfnamefont
  {E.~M.}\ \bibnamefont {Brauchle}}, \bibinfo {author} {\bibfnamefont {H.~J.}\
  \bibnamefont {Byrne}}, \bibinfo {author} {\bibfnamefont {I.}~\bibnamefont
  {Chourpa}}, \bibinfo {author} {\bibfnamefont {R.}~\bibnamefont {Cicchi}},
  \bibinfo {author} {\bibfnamefont {F.}~\bibnamefont {Cuisinier}}, \bibinfo
  {author} {\bibfnamefont {M.}~\bibnamefont {Culha}}, \bibinfo {author}
  {\bibfnamefont {M.}~\bibnamefont {Dahms}}, \bibinfo {author} {\bibfnamefont
  {C.}~\bibnamefont {David}}, \bibinfo {author} {\bibfnamefont
  {L.}~\bibnamefont {Duponchel}}, \bibinfo {author} {\bibfnamefont
  {S.}~\bibnamefont {Duraipandian}}, \bibinfo {author} {\bibfnamefont {S.~F.}\
  \bibnamefont {El-Mashtoly}}, \bibinfo {author} {\bibfnamefont {D.~I.}\
  \bibnamefont {Ellis}}, \bibinfo {author} {\bibfnamefont {G.}~\bibnamefont
  {Eppe}}, \bibinfo {author} {\bibfnamefont {G.}~\bibnamefont {Falgayrac}},
  \bibinfo {author} {\bibfnamefont {O.}~\bibnamefont {Gamulin}}, \bibinfo
  {author} {\bibfnamefont {B.}~\bibnamefont {Gardner}}, \bibinfo {author}
  {\bibfnamefont {P.}~\bibnamefont {Gardner}}, \bibinfo {author} {\bibfnamefont
  {K.}~\bibnamefont {Gerwert}}, \bibinfo {author} {\bibfnamefont {E.~J.}\
  \bibnamefont {Giamarellos-Bourboulis}}, \bibinfo {author} {\bibfnamefont
  {S.}~\bibnamefont {Gizurarson}}, \bibinfo {author} {\bibfnamefont
  {M.}~\bibnamefont {Gnyba}}, \bibinfo {author} {\bibfnamefont
  {R.}~\bibnamefont {Goodacre}}, \bibinfo {author} {\bibfnamefont
  {P.}~\bibnamefont {Grysan}}, \bibinfo {author} {\bibfnamefont
  {O.}~\bibnamefont {Guntinas-Lichius}}, \bibinfo {author} {\bibfnamefont
  {H.}~\bibnamefont {Helgadottir}}, \bibinfo {author} {\bibfnamefont {V.~M.}\
  \bibnamefont {Gro{\v{s}}ev}}, \bibinfo {author} {\bibfnamefont
  {C.}~\bibnamefont {Kendall}}, \bibinfo {author} {\bibfnamefont
  {R.}~\bibnamefont {Kiselev}}, \bibinfo {author} {\bibfnamefont
  {M.}~\bibnamefont {K{\"{o}}lbach}}, \bibinfo {author} {\bibfnamefont
  {C.}~\bibnamefont {Krafft}}, \bibinfo {author} {\bibfnamefont
  {S.}~\bibnamefont {Krishnamoorthy}}, \bibinfo {author} {\bibfnamefont
  {P.}~\bibnamefont {Kubryck}}, \bibinfo {author} {\bibfnamefont
  {B.}~\bibnamefont {Lendl}}, \bibinfo {author} {\bibfnamefont
  {P.}~\bibnamefont {Loza-Alvarez}}, \bibinfo {author} {\bibfnamefont {F.~M.}\
  \bibnamefont {Lyng}}, \bibinfo {author} {\bibfnamefont {S.}~\bibnamefont
  {Machill}}, \bibinfo {author} {\bibfnamefont {C.}~\bibnamefont {Malherbe}},
  \bibinfo {author} {\bibfnamefont {M.}~\bibnamefont {Marro}}, \bibinfo
  {author} {\bibfnamefont {M.~P.~M.}\ \bibnamefont {Marques}}, \bibinfo
  {author} {\bibfnamefont {E.}~\bibnamefont {Matuszyk}}, \bibinfo {author}
  {\bibfnamefont {C.~F.}\ \bibnamefont {Morasso}}, \bibinfo {author}
  {\bibfnamefont {M.}~\bibnamefont {Moreau}}, \bibinfo {author} {\bibfnamefont
  {H.}~\bibnamefont {Muhamadali}}, \bibinfo {author} {\bibfnamefont
  {V.}~\bibnamefont {Mussi}}, \bibinfo {author} {\bibfnamefont
  {I.}~\bibnamefont {Notingher}}, \bibinfo {author} {\bibfnamefont {M.~Z.}\
  \bibnamefont {Pacia}}, \bibinfo {author} {\bibfnamefont {F.~S.}\ \bibnamefont
  {Pavone}}, \bibinfo {author} {\bibfnamefont {G.}~\bibnamefont {Penel}},
  \bibinfo {author} {\bibfnamefont {D.}~\bibnamefont {Petersen}}, \bibinfo
  {author} {\bibfnamefont {O.}~\bibnamefont {Piot}}, \bibinfo {author}
  {\bibfnamefont {J.~V.}\ \bibnamefont {Rau}}, \bibinfo {author} {\bibfnamefont
  {M.}~\bibnamefont {Richter}}, \bibinfo {author} {\bibfnamefont {M.~K.}\
  \bibnamefont {Rybarczyk}}, \bibinfo {author} {\bibfnamefont {H.}~\bibnamefont
  {Salehi}}, \bibinfo {author} {\bibfnamefont {K.}~\bibnamefont
  {Schenke-Layland}}, \bibinfo {author} {\bibfnamefont {S.}~\bibnamefont
  {Schl{\"{u}}cker}}, \bibinfo {author} {\bibfnamefont {M.}~\bibnamefont
  {Schosserer}}, \bibinfo {author} {\bibfnamefont {K.}~\bibnamefont
  {Sch{\"{u}}tze}}, \bibinfo {author} {\bibfnamefont {V.}~\bibnamefont
  {Sergo}}, \bibinfo {author} {\bibfnamefont {F.}~\bibnamefont {Sinjab}},
  \bibinfo {author} {\bibfnamefont {J.}~\bibnamefont {Smulko}}, \bibinfo
  {author} {\bibfnamefont {G.~D.}\ \bibnamefont {Sockalingum}}, \bibinfo
  {author} {\bibfnamefont {C.}~\bibnamefont {Stiebing}}, \bibinfo {author}
  {\bibfnamefont {N.}~\bibnamefont {Stone}}, \bibinfo {author} {\bibfnamefont
  {V.}~\bibnamefont {Untereiner}}, \bibinfo {author} {\bibfnamefont
  {R.}~\bibnamefont {Vanna}}, \bibinfo {author} {\bibfnamefont
  {K.}~\bibnamefont {Wieland}}, \bibinfo {author} {\bibfnamefont
  {J.}~\bibnamefont {Popp}},\ and\ \bibinfo {author} {\bibfnamefont
  {T.}~\bibnamefont {Bocklitz}},\ }\bibfield  {title} {\bibinfo {title}
  {{Comparability of Raman Spectroscopic Configurations: A Large Scale
  Cross-Laboratory Study}},\ }\href
  {https://doi.org/10.1021/acs.analchem.0c02696} {\bibfield  {journal}
  {\bibinfo  {journal} {Analytical Chemistry}\ }\textbf {\bibinfo {volume}
  {92}},\ \bibinfo {pages} {15745} (\bibinfo {year} {2020})}\BibitemShut
  {NoStop}%
\bibitem [{\citenamefont {Powell}\ \emph {et~al.}(1982)\citenamefont {Powell},
  \citenamefont {Erickson},\ and\ \citenamefont {Madey}}]{Powell1982}%
  \BibitemOpen
  \bibfield  {author} {\bibinfo {author} {\bibfnamefont {C.~J.}\ \bibnamefont
  {Powell}}, \bibinfo {author} {\bibfnamefont {N.~E.}\ \bibnamefont
  {Erickson}},\ and\ \bibinfo {author} {\bibfnamefont {T.~E.}\ \bibnamefont
  {Madey}},\ }\bibfield  {title} {\bibinfo {title} {{Results of a joint
  auger/esca round robin sponsored by astm committee E-42 on surface analysis.
  Part II. Auger results}},\ }\href
  {https://doi.org/https://doi.org/10.1016/0368-2048(82)85010-X} {\bibfield
  {journal} {\bibinfo  {journal} {Journal of Electron Spectroscopy and Related
  Phenomena}\ }\textbf {\bibinfo {volume} {25}},\ \bibinfo {pages} {87}
  (\bibinfo {year} {1982})}\BibitemShut {NoStop}%
\bibitem [{\citenamefont {Bristow}\ and\ \citenamefont
  {Webb}(2003)}]{Bristow2003}%
  \BibitemOpen
  \bibfield  {author} {\bibinfo {author} {\bibfnamefont {A.~W.~T.}\
  \bibnamefont {Bristow}}\ and\ \bibinfo {author} {\bibfnamefont {K.~S.}\
  \bibnamefont {Webb}},\ }\bibfield  {title} {\bibinfo {title}
  {{Intercomparison study on accurate mass measurement of small molecules in
  mass spectrometry}},\ }\href {https://doi.org/10.1016/S1044-0305(03)00403-3}
  {\bibfield  {journal} {\bibinfo  {journal} {Journal of the American Society
  for Mass Spectrometry}\ }\textbf {\bibinfo {volume} {14}},\ \bibinfo {pages}
  {1086} (\bibinfo {year} {2003})}\BibitemShut {NoStop}%
\bibitem [{\citenamefont {Leymarie}\ \emph {et~al.}(2013)\citenamefont
  {Leymarie}, \citenamefont {Griffin}, \citenamefont {Jonscher}, \citenamefont
  {Kolarich}, \citenamefont {Orlando}, \citenamefont {McComb}, \citenamefont
  {Zaia}, \citenamefont {Aguilan}, \citenamefont {Alley}, \citenamefont
  {Altmann}, \citenamefont {Ball}, \citenamefont {Basumallick}, \citenamefont
  {Bazemore-Walker}, \citenamefont {Behnken}, \citenamefont {Blank},
  \citenamefont {Brown}, \citenamefont {Bunz}, \citenamefont {Cairo},
  \citenamefont {Cipollo}, \citenamefont {Daneshfar}, \citenamefont {Desaire},
  \citenamefont {Drake}, \citenamefont {Go}, \citenamefont {Goldman},
  \citenamefont {Gruber}, \citenamefont {Halim}, \citenamefont {Hathout},
  \citenamefont {Hensbergen}, \citenamefont {Horn}, \citenamefont {Hurum},
  \citenamefont {Jabs}, \citenamefont {Larson}, \citenamefont {Ly},
  \citenamefont {Mann}, \citenamefont {Marx}, \citenamefont {Mechref},
  \citenamefont {Meyer}, \citenamefont {M{\"{o}}ginger}, \citenamefont
  {Neus{\"{u}}{\ss}}, \citenamefont {Nilsson}, \citenamefont {Novotny},
  \citenamefont {Nyalwidhe}, \citenamefont {Packer}, \citenamefont {Pompach},
  \citenamefont {Reiz}, \citenamefont {Resemann}, \citenamefont {Rohrer},
  \citenamefont {Ruthenbeck}, \citenamefont {Sanda}, \citenamefont {Schulz},
  \citenamefont {Schweiger-Hufnagel}, \citenamefont {Sihlbom}, \citenamefont
  {Song}, \citenamefont {Staples}, \citenamefont {Suckau}, \citenamefont
  {Tang}, \citenamefont {Thaysen-Andersen}, \citenamefont {Viner},
  \citenamefont {An}, \citenamefont {Valmu}, \citenamefont {Wada},
  \citenamefont {Watson}, \citenamefont {Windwarder}, \citenamefont {Whittal},
  \citenamefont {Wuhrer}, \citenamefont {Zhu},\ and\ \citenamefont
  {Zou}}]{Leymarie2013}%
  \BibitemOpen
  \bibfield  {author} {\bibinfo {author} {\bibfnamefont {N.}~\bibnamefont
  {Leymarie}}, \bibinfo {author} {\bibfnamefont {P.~J.}\ \bibnamefont
  {Griffin}}, \bibinfo {author} {\bibfnamefont {K.}~\bibnamefont {Jonscher}},
  \bibinfo {author} {\bibfnamefont {D.}~\bibnamefont {Kolarich}}, \bibinfo
  {author} {\bibfnamefont {R.}~\bibnamefont {Orlando}}, \bibinfo {author}
  {\bibfnamefont {M.}~\bibnamefont {McComb}}, \bibinfo {author} {\bibfnamefont
  {J.}~\bibnamefont {Zaia}}, \bibinfo {author} {\bibfnamefont {J.}~\bibnamefont
  {Aguilan}}, \bibinfo {author} {\bibfnamefont {W.~R.}\ \bibnamefont {Alley}},
  \bibinfo {author} {\bibfnamefont {F.}~\bibnamefont {Altmann}}, \bibinfo
  {author} {\bibfnamefont {L.~E.}\ \bibnamefont {Ball}}, \bibinfo {author}
  {\bibfnamefont {L.}~\bibnamefont {Basumallick}}, \bibinfo {author}
  {\bibfnamefont {C.~R.}\ \bibnamefont {Bazemore-Walker}}, \bibinfo {author}
  {\bibfnamefont {H.}~\bibnamefont {Behnken}}, \bibinfo {author} {\bibfnamefont
  {M.~A.}\ \bibnamefont {Blank}}, \bibinfo {author} {\bibfnamefont {K.~J.}\
  \bibnamefont {Brown}}, \bibinfo {author} {\bibfnamefont {S.~C.}\ \bibnamefont
  {Bunz}}, \bibinfo {author} {\bibfnamefont {C.~W.}\ \bibnamefont {Cairo}},
  \bibinfo {author} {\bibfnamefont {J.~F.}\ \bibnamefont {Cipollo}}, \bibinfo
  {author} {\bibfnamefont {R.}~\bibnamefont {Daneshfar}}, \bibinfo {author}
  {\bibfnamefont {H.}~\bibnamefont {Desaire}}, \bibinfo {author} {\bibfnamefont
  {R.~R.}\ \bibnamefont {Drake}}, \bibinfo {author} {\bibfnamefont {E.~P.}\
  \bibnamefont {Go}}, \bibinfo {author} {\bibfnamefont {R.}~\bibnamefont
  {Goldman}}, \bibinfo {author} {\bibfnamefont {C.}~\bibnamefont {Gruber}},
  \bibinfo {author} {\bibfnamefont {A.}~\bibnamefont {Halim}}, \bibinfo
  {author} {\bibfnamefont {Y.}~\bibnamefont {Hathout}}, \bibinfo {author}
  {\bibfnamefont {P.~J.}\ \bibnamefont {Hensbergen}}, \bibinfo {author}
  {\bibfnamefont {D.~M.}\ \bibnamefont {Horn}}, \bibinfo {author}
  {\bibfnamefont {D.}~\bibnamefont {Hurum}}, \bibinfo {author} {\bibfnamefont
  {W.}~\bibnamefont {Jabs}}, \bibinfo {author} {\bibfnamefont {G.}~\bibnamefont
  {Larson}}, \bibinfo {author} {\bibfnamefont {M.}~\bibnamefont {Ly}}, \bibinfo
  {author} {\bibfnamefont {B.~F.}\ \bibnamefont {Mann}}, \bibinfo {author}
  {\bibfnamefont {K.}~\bibnamefont {Marx}}, \bibinfo {author} {\bibfnamefont
  {Y.}~\bibnamefont {Mechref}}, \bibinfo {author} {\bibfnamefont
  {B.}~\bibnamefont {Meyer}}, \bibinfo {author} {\bibfnamefont
  {U.}~\bibnamefont {M{\"{o}}ginger}}, \bibinfo {author} {\bibfnamefont
  {C.}~\bibnamefont {Neus{\"{u}}{\ss}}}, \bibinfo {author} {\bibfnamefont
  {J.}~\bibnamefont {Nilsson}}, \bibinfo {author} {\bibfnamefont {M.~V.}\
  \bibnamefont {Novotny}}, \bibinfo {author} {\bibfnamefont {J.~O.}\
  \bibnamefont {Nyalwidhe}}, \bibinfo {author} {\bibfnamefont {N.~H.}\
  \bibnamefont {Packer}}, \bibinfo {author} {\bibfnamefont {P.}~\bibnamefont
  {Pompach}}, \bibinfo {author} {\bibfnamefont {B.}~\bibnamefont {Reiz}},
  \bibinfo {author} {\bibfnamefont {A.}~\bibnamefont {Resemann}}, \bibinfo
  {author} {\bibfnamefont {J.~S.}\ \bibnamefont {Rohrer}}, \bibinfo {author}
  {\bibfnamefont {A.}~\bibnamefont {Ruthenbeck}}, \bibinfo {author}
  {\bibfnamefont {M.}~\bibnamefont {Sanda}}, \bibinfo {author} {\bibfnamefont
  {J.~M.}\ \bibnamefont {Schulz}}, \bibinfo {author} {\bibfnamefont
  {U.}~\bibnamefont {Schweiger-Hufnagel}}, \bibinfo {author} {\bibfnamefont
  {C.}~\bibnamefont {Sihlbom}}, \bibinfo {author} {\bibfnamefont
  {E.}~\bibnamefont {Song}}, \bibinfo {author} {\bibfnamefont {G.~O.}\
  \bibnamefont {Staples}}, \bibinfo {author} {\bibfnamefont {D.}~\bibnamefont
  {Suckau}}, \bibinfo {author} {\bibfnamefont {H.}~\bibnamefont {Tang}},
  \bibinfo {author} {\bibfnamefont {M.}~\bibnamefont {Thaysen-Andersen}},
  \bibinfo {author} {\bibfnamefont {R.~I.}\ \bibnamefont {Viner}}, \bibinfo
  {author} {\bibfnamefont {Y.}~\bibnamefont {An}}, \bibinfo {author}
  {\bibfnamefont {L.}~\bibnamefont {Valmu}}, \bibinfo {author} {\bibfnamefont
  {Y.}~\bibnamefont {Wada}}, \bibinfo {author} {\bibfnamefont {M.}~\bibnamefont
  {Watson}}, \bibinfo {author} {\bibfnamefont {M.}~\bibnamefont {Windwarder}},
  \bibinfo {author} {\bibfnamefont {R.}~\bibnamefont {Whittal}}, \bibinfo
  {author} {\bibfnamefont {M.}~\bibnamefont {Wuhrer}}, \bibinfo {author}
  {\bibfnamefont {Y.}~\bibnamefont {Zhu}},\ and\ \bibinfo {author}
  {\bibfnamefont {C.}~\bibnamefont {Zou}},\ }\bibfield  {title} {\bibinfo
  {title} {{Interlaboratory study on differential analysis of protein
  glycosylation by mass spectrometry: The ABRF glycoprotein research
  multi-institutional study 2012}},\ }\href
  {https://doi.org/10.1074/mcp.M113.030643} {\bibfield  {journal} {\bibinfo
  {journal} {Molecular and Cellular Proteomics}\ }\textbf {\bibinfo {volume}
  {12}},\ \bibinfo {pages} {2935} (\bibinfo {year} {2013})}\BibitemShut
  {NoStop}%
\bibitem [{\citenamefont {Blum}\ \emph {et~al.}(2014)\citenamefont {Blum},
  \citenamefont {Opilik}, \citenamefont {Atkin}, \citenamefont {Braun},
  \citenamefont {K{\"{a}}mmer}, \citenamefont {Kravtsov}, \citenamefont
  {Kumar}, \citenamefont {Lemeshko}, \citenamefont {Li}, \citenamefont
  {Luszcz}, \citenamefont {Maleki}, \citenamefont {Meixner}, \citenamefont
  {Minne}, \citenamefont {Raschke}, \citenamefont {Ren}, \citenamefont
  {Rogalski}, \citenamefont {Roy}, \citenamefont {Stephanidis}, \citenamefont
  {Wang}, \citenamefont {Zhang}, \citenamefont {Zhong},\ and\ \citenamefont
  {Zenobi}}]{Blum2013}%
  \BibitemOpen
  \bibfield  {author} {\bibinfo {author} {\bibfnamefont {C.}~\bibnamefont
  {Blum}}, \bibinfo {author} {\bibfnamefont {L.}~\bibnamefont {Opilik}},
  \bibinfo {author} {\bibfnamefont {J.~M.}\ \bibnamefont {Atkin}}, \bibinfo
  {author} {\bibfnamefont {K.}~\bibnamefont {Braun}}, \bibinfo {author}
  {\bibfnamefont {S.~B.}\ \bibnamefont {K{\"{a}}mmer}}, \bibinfo {author}
  {\bibfnamefont {V.}~\bibnamefont {Kravtsov}}, \bibinfo {author}
  {\bibfnamefont {N.}~\bibnamefont {Kumar}}, \bibinfo {author} {\bibfnamefont
  {S.}~\bibnamefont {Lemeshko}}, \bibinfo {author} {\bibfnamefont {J.-F.}\
  \bibnamefont {Li}}, \bibinfo {author} {\bibfnamefont {K.}~\bibnamefont
  {Luszcz}}, \bibinfo {author} {\bibfnamefont {T.}~\bibnamefont {Maleki}},
  \bibinfo {author} {\bibfnamefont {A.~J.}\ \bibnamefont {Meixner}}, \bibinfo
  {author} {\bibfnamefont {S.}~\bibnamefont {Minne}}, \bibinfo {author}
  {\bibfnamefont {M.~B.}\ \bibnamefont {Raschke}}, \bibinfo {author}
  {\bibfnamefont {B.}~\bibnamefont {Ren}}, \bibinfo {author} {\bibfnamefont
  {J.}~\bibnamefont {Rogalski}}, \bibinfo {author} {\bibfnamefont
  {D.}~\bibnamefont {Roy}}, \bibinfo {author} {\bibfnamefont {B.}~\bibnamefont
  {Stephanidis}}, \bibinfo {author} {\bibfnamefont {X.}~\bibnamefont {Wang}},
  \bibinfo {author} {\bibfnamefont {D.}~\bibnamefont {Zhang}}, \bibinfo
  {author} {\bibfnamefont {J.-H.}\ \bibnamefont {Zhong}},\ and\ \bibinfo
  {author} {\bibfnamefont {R.}~\bibnamefont {Zenobi}},\ }\bibfield  {title}
  {\bibinfo {title} {{Tip-enhanced Raman spectroscopy – an interlaboratory
  reproducibility and comparison study}},\ }\href
  {https://doi.org/https://doi.org/10.1002/jrs.4423} {\bibfield  {journal}
  {\bibinfo  {journal} {Journal of Raman Spectroscopy}\ }\textbf {\bibinfo
  {volume} {45}},\ \bibinfo {pages} {22} (\bibinfo {year} {2014})}\BibitemShut
  {NoStop}%
\bibitem [{\citenamefont {Lynch}\ \emph {et~al.}(1976)\citenamefont {Lynch},
  \citenamefont {Kramer}, \citenamefont {Lotem},\ and\ \citenamefont
  {Bloembergen}}]{lotem1976}%
  \BibitemOpen
  \bibfield  {author} {\bibinfo {author} {\bibfnamefont {R.~T.}\ \bibnamefont
  {Lynch}}, \bibinfo {author} {\bibfnamefont {S.~D.}\ \bibnamefont {Kramer}},
  \bibinfo {author} {\bibfnamefont {H.}~\bibnamefont {Lotem}},\ and\ \bibinfo
  {author} {\bibfnamefont {N.}~\bibnamefont {Bloembergen}},\ }\bibfield
  {title} {\bibinfo {title} {{Double resonance interference in third-order
  light mixing}},\ }\href
  {https://doi.org/https://doi.org/10.1016/0030-4018(76)90020-1} {\bibfield
  {journal} {\bibinfo  {journal} {Optics Communications}\ }\textbf {\bibinfo
  {volume} {16}},\ \bibinfo {pages} {372} (\bibinfo {year} {1976})}\BibitemShut
  {NoStop}%
\bibitem [{\citenamefont {Vernuccio}\ \emph {et~al.}(2022)\citenamefont
  {Vernuccio}, \citenamefont {Bresci}, \citenamefont {Talone}, \citenamefont
  {de~la Cadena}, \citenamefont {Ceconello}, \citenamefont {Mantero},
  \citenamefont {Sobacchi}, \citenamefont {Vanna}, \citenamefont {Cerullo},\
  and\ \citenamefont {Polli}}]{Vernuccio:22}%
  \BibitemOpen
  \bibfield  {author} {\bibinfo {author} {\bibfnamefont {F.}~\bibnamefont
  {Vernuccio}}, \bibinfo {author} {\bibfnamefont {A.}~\bibnamefont {Bresci}},
  \bibinfo {author} {\bibfnamefont {B.}~\bibnamefont {Talone}}, \bibinfo
  {author} {\bibfnamefont {A.}~\bibnamefont {de~la Cadena}}, \bibinfo {author}
  {\bibfnamefont {C.}~\bibnamefont {Ceconello}}, \bibinfo {author}
  {\bibfnamefont {S.}~\bibnamefont {Mantero}}, \bibinfo {author} {\bibfnamefont
  {C.}~\bibnamefont {Sobacchi}}, \bibinfo {author} {\bibfnamefont
  {R.}~\bibnamefont {Vanna}}, \bibinfo {author} {\bibfnamefont
  {G.}~\bibnamefont {Cerullo}},\ and\ \bibinfo {author} {\bibfnamefont
  {D.}~\bibnamefont {Polli}},\ }\bibfield  {title} {\bibinfo {title}
  {{Fingerprint multiplex CARS at high speed based on supercontinuum generation
  in bulk media and deep learning spectral denoising}},\ }\href
  {https://doi.org/10.1364/OE.463032} {\bibfield  {journal} {\bibinfo
  {journal} {Opt. Express}\ }\textbf {\bibinfo {volume} {30}},\ \bibinfo
  {pages} {30135} (\bibinfo {year} {2022})}\BibitemShut {NoStop}%
\bibitem [{\citenamefont {Gra{\v{z}}ulevi{\v{c}}iūtė}\ \emph
  {et~al.}(2015)\citenamefont {Gra{\v{z}}ulevi{\v{c}}iūtė}, \citenamefont
  {Skeivytė}, \citenamefont {Keblytė}, \citenamefont {Galinis}, \citenamefont
  {Tamo{\v{s}}auskas},\ and\ \citenamefont {Dubietis}}]{Grazuleviciute2015}%
  \BibitemOpen
  \bibfield  {author} {\bibinfo {author} {\bibfnamefont {I.}~\bibnamefont
  {Gra{\v{z}}ulevi{\v{c}}iūtė}}, \bibinfo {author} {\bibfnamefont
  {M.}~\bibnamefont {Skeivytė}}, \bibinfo {author} {\bibfnamefont
  {E.}~\bibnamefont {Keblytė}}, \bibinfo {author} {\bibfnamefont
  {J.}~\bibnamefont {Galinis}}, \bibinfo {author} {\bibfnamefont
  {G.}~\bibnamefont {Tamo{\v{s}}auskas}},\ and\ \bibinfo {author}
  {\bibfnamefont {A.}~\bibnamefont {Dubietis}},\ }\bibfield  {title} {\bibinfo
  {title} {{Supercontinuum generation in YAG and sapphire with picosecond laser
  pulses}},\ }\href {https://doi.org/10.3952/physics.v55i2.3101} {\bibfield
  {journal} {\bibinfo  {journal} {Lithuanian Journal of Physics}\ }\textbf
  {\bibinfo {volume} {55}},\ \bibinfo {pages} {110} (\bibinfo {year}
  {2015})}\BibitemShut {NoStop}%
\bibitem [{\citenamefont {Krivovichev}(2014)}]{Krivovichev2014}%
  \BibitemOpen
  \bibfield  {author} {\bibinfo {author} {\bibfnamefont {S.~V.}\ \bibnamefont
  {Krivovichev}},\ }\bibfield  {title} {\bibinfo {title} {{Which inorganic
  structures are the most complex?}},\ }\href
  {https://doi.org/10.1002/anie.201304374} {\bibfield  {journal} {\bibinfo
  {journal} {Angewandte Chemie - International Edition}\ }\textbf {\bibinfo
  {volume} {53}},\ \bibinfo {pages} {654} (\bibinfo {year} {2014})}\BibitemShut
  {NoStop}%
\bibitem [{\citenamefont {Kau{\ss}ler}\ and\ \citenamefont
  {Kieslich}(2021)}]{Kaussler:oc5005}%
  \BibitemOpen
  \bibfield  {author} {\bibinfo {author} {\bibfnamefont {C.}~\bibnamefont
  {Kau{\ss}ler}}\ and\ \bibinfo {author} {\bibfnamefont {G.}~\bibnamefont
  {Kieslich}},\ }\bibfield  {title} {\bibinfo {title} {{{\it crystIT}:
  complexity and configurational entropy of crystal structures via information
  theory}},\ }\href {https://doi.org/10.1107/S1600576720016386} {\bibfield
  {journal} {\bibinfo  {journal} {Journal of Applied Crystallography}\ }\textbf
  {\bibinfo {volume} {54}},\ \bibinfo {pages} {306} (\bibinfo {year}
  {2021})}\BibitemShut {NoStop}%
\bibitem [{\citenamefont
  {Villars}(2016{\natexlab{a}})}]{Villars2016:sm_isp_sd_1500919}%
  \BibitemOpen
  \bibfield  {author} {\bibinfo {author} {\bibfnamefont {P.}~\bibnamefont
  {Villars}},\ }\href
  {https://materials.springer.com/isp/crystallographic/docs/sd_1500919}
  {\bibinfo {title} {{C, diamond (C dia) Crystal Structure: Datasheet from
  ``PAULING FILE Multinaries Edition -- 2012'' in SpringerMaterials}}}
  (\bibinfo {year} {2016}{\natexlab{a}})\BibitemShut {NoStop}%
\bibitem [{\citenamefont {Krishnan}(1944)}]{Krishnan1944}%
  \BibitemOpen
  \bibfield  {author} {\bibinfo {author} {\bibfnamefont {R.~S.}\ \bibnamefont
  {Krishnan}},\ }\bibfield  {title} {\bibinfo {title} {{The raman spectrum of
  diamond}},\ }\href {https://doi.org/10.1007/BF03173448} {\bibfield  {journal}
  {\bibinfo  {journal} {Proceedings of the Indian Academy of Sciences - Section
  A}\ }\textbf {\bibinfo {volume} {19}},\ \bibinfo {pages} {216} (\bibinfo
  {year} {1944})}\BibitemShut {NoStop}%
\bibitem [{\citenamefont
  {Villars}(2016{\natexlab{b}})}]{Villars2016:sm_isp_sd_1628877}%
  \BibitemOpen
  \bibfield  {author} {\bibinfo {author} {\bibfnamefont {P.}~\bibnamefont
  {Villars}},\ }\href
  {https://materials.springer.com/isp/crystallographic/docs/sd_1628877}
  {\bibinfo {title} {{4H-SiC (SiC 4H) Crystal Structure: Datasheet from
  ``PAULING FILE Multinaries Edition -- 2012'' in SpringerMaterials}}}
  (\bibinfo {year} {2016}{\natexlab{b}})\BibitemShut {NoStop}%
\bibitem [{\citenamefont {Feldman}\ \emph {et~al.}(1968)\citenamefont
  {Feldman}, \citenamefont {Parker}, \citenamefont {Choyke},\ and\
  \citenamefont {Patrick}}]{Feldman1968}%
  \BibitemOpen
  \bibfield  {author} {\bibinfo {author} {\bibfnamefont {D.~W.}\ \bibnamefont
  {Feldman}}, \bibinfo {author} {\bibfnamefont {J.~H.}\ \bibnamefont {Parker}},
  \bibinfo {author} {\bibfnamefont {W.~J.}\ \bibnamefont {Choyke}},\ and\
  \bibinfo {author} {\bibfnamefont {L.}~\bibnamefont {Patrick}},\ }\bibfield
  {title} {\bibinfo {title} {{Phonon dispersion curves by raman scattering in
  SiC, polytypes 3C, 4H, 6H, 15R, and 21R}},\ }\href
  {https://doi.org/10.1103/PhysRev.173.787} {\bibfield  {journal} {\bibinfo
  {journal} {Physical Review}\ }\textbf {\bibinfo {volume} {173}},\ \bibinfo
  {pages} {787} (\bibinfo {year} {1968})}\BibitemShut {NoStop}%
\bibitem [{\citenamefont
  {Villars}(2016{\natexlab{c}})}]{Villars2016:sm_isp_sd_1145237}%
  \BibitemOpen
  \bibfield  {author} {\bibinfo {author} {\bibfnamefont {P.}~\bibnamefont
  {Villars}},\ }\href
  {https://materials.springer.com/isp/crystallographic/docs/sd_1145237}
  {\bibinfo {title} {{KDP/TiO2-1 (Pr-1) (H2K[PO4] rt) Crystal Structure:
  Datasheet from ``PAULING FILE Multinaries Edition -- 2012'' in
  SpringerMaterials}}} (\bibinfo {year} {2016}{\natexlab{c}})\BibitemShut
  {NoStop}%
\bibitem [{\citenamefont {Lu}\ and\ \citenamefont {Sun}(2002)}]{Lu2002}%
  \BibitemOpen
  \bibfield  {author} {\bibinfo {author} {\bibfnamefont {G.~W.}\ \bibnamefont
  {Lu}}\ and\ \bibinfo {author} {\bibfnamefont {X.}~\bibnamefont {Sun}},\
  }\bibfield  {title} {\bibinfo {title} {{Raman study of lattice vibration
  modes and growth mechanism of KDP single crystals}},\ }\href
  {https://doi.org/10.1002/1521-4079(200202)37:1<93::AID-CRAT93>3.0.CO;2-3}
  {\bibfield  {journal} {\bibinfo  {journal} {Crystal Research and Technology}\
  }\textbf {\bibinfo {volume} {37}},\ \bibinfo {pages} {93} (\bibinfo {year}
  {2002})}\BibitemShut {NoStop}%
\bibitem [{\citenamefont {Allan}\ and\ \citenamefont
  {Nelmes}(1996)}]{Allan1996}%
  \BibitemOpen
  \bibfield  {author} {\bibinfo {author} {\bibfnamefont {D.~R.}\ \bibnamefont
  {Allan}}\ and\ \bibinfo {author} {\bibfnamefont {R.~J.}\ \bibnamefont
  {Nelmes}},\ }\bibfield  {title} {\bibinfo {title} {{The structural pressure
  dependence of potassium titanyl phosphate (KTP) to 8 GPa}},\ }\href
  {https://doi.org/10.1088/0953-8984/8/14/010} {\bibfield  {journal} {\bibinfo
  {journal} {Journal of Physics Condensed Matter}\ }\textbf {\bibinfo {volume}
  {8}},\ \bibinfo {pages} {2337} (\bibinfo {year} {1996})}\BibitemShut
  {NoStop}%
\bibitem [{\citenamefont {Cicerone}\ \emph {et~al.}(2012)\citenamefont
  {Cicerone}, \citenamefont {Aamer}, \citenamefont {Lee},\ and\ \citenamefont
  {Vartiainen}}]{Cicerone2012}%
  \BibitemOpen
  \bibfield  {author} {\bibinfo {author} {\bibfnamefont {M.~T.}\ \bibnamefont
  {Cicerone}}, \bibinfo {author} {\bibfnamefont {K.~A.}\ \bibnamefont {Aamer}},
  \bibinfo {author} {\bibfnamefont {Y.~J.}\ \bibnamefont {Lee}},\ and\ \bibinfo
  {author} {\bibfnamefont {E.}~\bibnamefont {Vartiainen}},\ }\bibfield  {title}
  {\bibinfo {title} {{Maximum entropy and time-domain Kramers-Kronig phase
  retrieval approaches are functionally equivalent for CARS
  microspectroscopy}},\ }\href {https://doi.org/10.1002/jrs.3169} {\bibfield
  {journal} {\bibinfo  {journal} {Journal of Raman Spectroscopy}\ }\textbf
  {\bibinfo {volume} {43}},\ \bibinfo {pages} {637} (\bibinfo {year}
  {2012})}\BibitemShut {NoStop}%
\bibitem [{\citenamefont {Burton}\ \emph {et~al.}(1998)\citenamefont {Burton},
  \citenamefont {Sun}, \citenamefont {Pophristic}, \citenamefont {Lukacs},
  \citenamefont {Long}, \citenamefont {Feng},\ and\ \citenamefont
  {Ferguson}}]{Burton1998}%
  \BibitemOpen
  \bibfield  {author} {\bibinfo {author} {\bibfnamefont {J.~C.}\ \bibnamefont
  {Burton}}, \bibinfo {author} {\bibfnamefont {L.}~\bibnamefont {Sun}},
  \bibinfo {author} {\bibfnamefont {M.}~\bibnamefont {Pophristic}}, \bibinfo
  {author} {\bibfnamefont {S.~J.}\ \bibnamefont {Lukacs}}, \bibinfo {author}
  {\bibfnamefont {F.~H.}\ \bibnamefont {Long}}, \bibinfo {author}
  {\bibfnamefont {Z.~C.}\ \bibnamefont {Feng}},\ and\ \bibinfo {author}
  {\bibfnamefont {I.~T.}\ \bibnamefont {Ferguson}},\ }\bibfield  {title}
  {\bibinfo {title} {{Spatial characterization of doped SiC wafers by Raman
  spectroscopy}},\ }\href {https://doi.org/10.1063/1.368947} {\bibfield
  {journal} {\bibinfo  {journal} {Journal of Applied Physics}\ }\textbf
  {\bibinfo {volume} {84}},\ \bibinfo {pages} {6268} (\bibinfo {year}
  {1998})}\BibitemShut {NoStop}%
\bibitem [{\citenamefont {Cheng}\ and\ \citenamefont {Xie}(2004)}]{Cheng2004}%
  \BibitemOpen
  \bibfield  {author} {\bibinfo {author} {\bibfnamefont {J.-X.}\ \bibnamefont
  {Cheng}}\ and\ \bibinfo {author} {\bibfnamefont {X.~S.}\ \bibnamefont
  {Xie}},\ }\bibfield  {title} {\bibinfo {title} {{Coherent Anti-Stokes Raman
  Scattering Microscopy: Instrumentation, Theory, and Applications}},\ }\href
  {https://doi.org/10.1021/jp035693v} {\bibfield  {journal} {\bibinfo
  {journal} {The Journal of Physical Chemistry B}\ }\textbf {\bibinfo {volume}
  {108}},\ \bibinfo {pages} {827} (\bibinfo {year} {2004})}\BibitemShut
  {NoStop}%
\bibitem [{\citenamefont {Potma}\ \emph {et~al.}(2000)\citenamefont {Potma},
  \citenamefont {de~Boeij},\ and\ \citenamefont {Wiersma}}]{potma2000}%
  \BibitemOpen
  \bibfield  {author} {\bibinfo {author} {\bibfnamefont {E.~O.}\ \bibnamefont
  {Potma}}, \bibinfo {author} {\bibfnamefont {W.~P.}\ \bibnamefont
  {de~Boeij}},\ and\ \bibinfo {author} {\bibfnamefont {D.~A.}\ \bibnamefont
  {Wiersma}},\ }\bibfield  {title} {\bibinfo {title} {{Nonlinear coherent
  four-wave mixing in optical microscopy}},\ }\href
  {https://doi.org/10.1364/JOSAB.17.001678} {\bibfield  {journal} {\bibinfo
  {journal} {J. Opt. Soc. Am. B}\ }\textbf {\bibinfo {volume} {17}},\ \bibinfo
  {pages} {1678} (\bibinfo {year} {2000})}\BibitemShut {NoStop}%
\bibitem [{\citenamefont {Amber}\ \emph {et~al.}(2021)\citenamefont {Amber},
  \citenamefont {Kirbus}, \citenamefont {Eng},\ and\ \citenamefont
  {R{\"{u}}sing}}]{Amber2021}%
  \BibitemOpen
  \bibfield  {author} {\bibinfo {author} {\bibfnamefont {Z.~H.}\ \bibnamefont
  {Amber}}, \bibinfo {author} {\bibfnamefont {B.}~\bibnamefont {Kirbus}},
  \bibinfo {author} {\bibfnamefont {L.~M.}\ \bibnamefont {Eng}},\ and\ \bibinfo
  {author} {\bibfnamefont {M.}~\bibnamefont {R{\"{u}}sing}},\ }\bibfield
  {title} {\bibinfo {title} {{Quantifying the coherent interaction length of
  second-harmonic microscopy in lithium niobate confined nanostructures}},\
  }\href {https://doi.org/10.1063/5.0058996} {\bibfield  {journal} {\bibinfo
  {journal} {Journal of Applied Physics}\ }\textbf {\bibinfo {volume} {130}},\
  \bibinfo {pages} {1} (\bibinfo {year} {2021})},\ \Eprint
  {https://arxiv.org/abs/2108.03397} {arXiv:2108.03397} \BibitemShut {NoStop}%
\bibitem [{\citenamefont {Amber}\ \emph {et~al.}(2022)\citenamefont {Amber},
  \citenamefont {Spychala}, \citenamefont {Eng},\ and\ \citenamefont
  {R{\"{u}}sing}}]{Amber2022}%
  \BibitemOpen
  \bibfield  {author} {\bibinfo {author} {\bibfnamefont {Z.~H.}\ \bibnamefont
  {Amber}}, \bibinfo {author} {\bibfnamefont {K.~J.}\ \bibnamefont {Spychala}},
  \bibinfo {author} {\bibfnamefont {L.~M.}\ \bibnamefont {Eng}},\ and\ \bibinfo
  {author} {\bibfnamefont {M.}~\bibnamefont {R{\"{u}}sing}},\ }\bibfield
  {title} {\bibinfo {title} {{Nonlinear optical interactions in focused beams
  and nanosized structures}},\ }\bibfield  {journal} {\bibinfo  {journal}
  {Journal of Applied Physics}\ }\textbf {\bibinfo {volume} {132}},\ \href
  {https://doi.org/10.1063/5.0125926} {10.1063/5.0125926} (\bibinfo {year}
  {2022})\BibitemShut {NoStop}%
\bibitem [{\citenamefont {Spychala}\ \emph {et~al.}(2023)\citenamefont
  {Spychala}, \citenamefont {Amber}, \citenamefont {Eng},\ and\ \citenamefont
  {Ruesing}}]{Spychala2023}%
  \BibitemOpen
  \bibfield  {author} {\bibinfo {author} {\bibfnamefont {K.~J.}\ \bibnamefont
  {Spychala}}, \bibinfo {author} {\bibfnamefont {Z.~H.}\ \bibnamefont {Amber}},
  \bibinfo {author} {\bibfnamefont {L.~M.}\ \bibnamefont {Eng}},\ and\ \bibinfo
  {author} {\bibfnamefont {M.}~\bibnamefont {Ruesing}},\ }\bibfield  {title}
  {\bibinfo {title} {{Modeling nonlinear optical interactions of focused beams
  in bulk crystals and thin films: A phenomenological approach}},\ }\href
  {https://doi.org/10.1063/5.0136252} {\bibfield  {journal} {\bibinfo
  {journal} {Journal of Applied Physics}\ }\textbf {\bibinfo {volume} {133}},\
  \bibinfo {pages} {123105} (\bibinfo {year} {2023})}\BibitemShut {NoStop}%
\bibitem [{\citenamefont {Knight}\ and\ \citenamefont
  {White}(1989)}]{Knight1989}%
  \BibitemOpen
  \bibfield  {author} {\bibinfo {author} {\bibfnamefont {D.~S.}\ \bibnamefont
  {Knight}}\ and\ \bibinfo {author} {\bibfnamefont {W.~B.}\ \bibnamefont
  {White}},\ }\bibfield  {title} {\bibinfo {title} {{Characterization of
  diamond films by Raman spectroscopy}},\ }\href
  {https://doi.org/10.1557/JMR.1989.0385} {\bibfield  {journal} {\bibinfo
  {journal} {Journal of Materials Research}\ }\textbf {\bibinfo {volume} {4}},\
  \bibinfo {pages} {385} (\bibinfo {year} {1989})}\BibitemShut {NoStop}%
\bibitem [{\citenamefont {Kugel}\ \emph {et~al.}(1988)\citenamefont {Kugel},
  \citenamefont {Brehat}, \citenamefont {Wyncke}, \citenamefont {Fontana},
  \citenamefont {Marnier}, \citenamefont {{Carabatos Nedelec}},\ and\
  \citenamefont {Mangin}}]{Kugel1988}%
  \BibitemOpen
  \bibfield  {author} {\bibinfo {author} {\bibfnamefont {G.~E.}\ \bibnamefont
  {Kugel}}, \bibinfo {author} {\bibfnamefont {F.}~\bibnamefont {Brehat}},
  \bibinfo {author} {\bibfnamefont {B.}~\bibnamefont {Wyncke}}, \bibinfo
  {author} {\bibfnamefont {M.~D.}\ \bibnamefont {Fontana}}, \bibinfo {author}
  {\bibfnamefont {G.}~\bibnamefont {Marnier}}, \bibinfo {author} {\bibfnamefont
  {C.}~\bibnamefont {{Carabatos Nedelec}}},\ and\ \bibinfo {author}
  {\bibfnamefont {J.}~\bibnamefont {Mangin}},\ }\bibfield  {title} {\bibinfo
  {title} {{The vibrational spectrum of a ktiopo4 single crystal studied by
  raman and infrared reflectivity spectroscopy}},\ }\href
  {https://doi.org/10.1088/0022-3719/21/32/011} {\bibfield  {journal} {\bibinfo
   {journal} {Journal of Physics C: Solid State Physics}\ }\textbf {\bibinfo
  {volume} {21}},\ \bibinfo {pages} {5565} (\bibinfo {year}
  {1988})}\BibitemShut {NoStop}%
\bibitem [{\citenamefont {Vivekanandan}\ \emph {et~al.}(1997)\citenamefont
  {Vivekanandan}, \citenamefont {Selvasekarapandian}, \citenamefont
  {Kolandaivel}, \citenamefont {Sebastian},\ and\ \citenamefont
  {Suma}}]{Vivekanandan1997}%
  \BibitemOpen
  \bibfield  {author} {\bibinfo {author} {\bibfnamefont {K.}~\bibnamefont
  {Vivekanandan}}, \bibinfo {author} {\bibfnamefont {S.}~\bibnamefont
  {Selvasekarapandian}}, \bibinfo {author} {\bibfnamefont {P.}~\bibnamefont
  {Kolandaivel}}, \bibinfo {author} {\bibfnamefont {M.~T.}\ \bibnamefont
  {Sebastian}},\ and\ \bibinfo {author} {\bibfnamefont {S.}~\bibnamefont
  {Suma}},\ }\bibfield  {title} {\bibinfo {title} {{Raman and FT-IR
  spectroscopic characterisation of flux grown KTiOPO4 and KRbTiOPO4 non-linear
  optical crystals}},\ }\href {https://doi.org/10.1016/S0254-0584(97)80165-4}
  {\bibfield  {journal} {\bibinfo  {journal} {Materials Chemistry and Physics}\
  }\textbf {\bibinfo {volume} {49}},\ \bibinfo {pages} {204} (\bibinfo {year}
  {1997})}\BibitemShut {NoStop}%
\bibitem [{\citenamefont {Zhao}\ \emph {et~al.}(2020)\citenamefont {Zhao},
  \citenamefont {R{\"{u}}sing}, \citenamefont {Javid}, \citenamefont {Ling},
  \citenamefont {Li}, \citenamefont {Lin},\ and\ \citenamefont
  {Mookherjea}}]{Zhao:20}%
  \BibitemOpen
  \bibfield  {author} {\bibinfo {author} {\bibfnamefont {J.}~\bibnamefont
  {Zhao}}, \bibinfo {author} {\bibfnamefont {M.}~\bibnamefont {R{\"{u}}sing}},
  \bibinfo {author} {\bibfnamefont {U.~A.}\ \bibnamefont {Javid}}, \bibinfo
  {author} {\bibfnamefont {J.}~\bibnamefont {Ling}}, \bibinfo {author}
  {\bibfnamefont {M.}~\bibnamefont {Li}}, \bibinfo {author} {\bibfnamefont
  {Q.}~\bibnamefont {Lin}},\ and\ \bibinfo {author} {\bibfnamefont
  {S.}~\bibnamefont {Mookherjea}},\ }\bibfield  {title} {\bibinfo {title}
  {{Shallow-etched thin-film lithium niobate waveguides for highly-efficient
  second-harmonic generation}},\ }\href {https://doi.org/10.1364/OE.395545}
  {\bibfield  {journal} {\bibinfo  {journal} {Opt. Express}\ }\textbf {\bibinfo
  {volume} {28}},\ \bibinfo {pages} {19669} (\bibinfo {year}
  {2020})}\BibitemShut {NoStop}%
\bibitem [{\citenamefont {Lu}\ \emph {et~al.}(2019)\citenamefont {Lu},
  \citenamefont {Surya}, \citenamefont {Liu}, \citenamefont {Bruch},
  \citenamefont {Gong}, \citenamefont {Xu},\ and\ \citenamefont
  {Tang}}]{Lu:19}%
  \BibitemOpen
  \bibfield  {author} {\bibinfo {author} {\bibfnamefont {J.}~\bibnamefont
  {Lu}}, \bibinfo {author} {\bibfnamefont {J.~B.}\ \bibnamefont {Surya}},
  \bibinfo {author} {\bibfnamefont {X.}~\bibnamefont {Liu}}, \bibinfo {author}
  {\bibfnamefont {A.~W.}\ \bibnamefont {Bruch}}, \bibinfo {author}
  {\bibfnamefont {Z.}~\bibnamefont {Gong}}, \bibinfo {author} {\bibfnamefont
  {Y.}~\bibnamefont {Xu}},\ and\ \bibinfo {author} {\bibfnamefont {H.~X.}\
  \bibnamefont {Tang}},\ }\bibfield  {title} {\bibinfo {title} {{Periodically
  poled thin-film lithium niobate microring resonators with a second-harmonic
  generation efficiency of 250,000\%/W}},\ }\href
  {https://doi.org/10.1364/OPTICA.6.001455} {\bibfield  {journal} {\bibinfo
  {journal} {Optica}\ }\textbf {\bibinfo {volume} {6}},\ \bibinfo {pages}
  {1455} (\bibinfo {year} {2019})},\ \Eprint {https://arxiv.org/abs/1911.00083}
  {arXiv:1911.00083} \BibitemShut {NoStop}%
\bibitem [{\citenamefont {R{\"{u}}sing}\ \emph {et~al.}(2016)\citenamefont
  {R{\"{u}}sing}, \citenamefont {Eigner}, \citenamefont {Mackwitz},
  \citenamefont {Berth}, \citenamefont {Silberhorn},\ and\ \citenamefont
  {Zrenner}}]{Rusing2016a}%
  \BibitemOpen
  \bibfield  {author} {\bibinfo {author} {\bibfnamefont {M.}~\bibnamefont
  {R{\"{u}}sing}}, \bibinfo {author} {\bibfnamefont {C.}~\bibnamefont
  {Eigner}}, \bibinfo {author} {\bibfnamefont {P.}~\bibnamefont {Mackwitz}},
  \bibinfo {author} {\bibfnamefont {G.}~\bibnamefont {Berth}}, \bibinfo
  {author} {\bibfnamefont {C.}~\bibnamefont {Silberhorn}},\ and\ \bibinfo
  {author} {\bibfnamefont {A.}~\bibnamefont {Zrenner}},\ }\bibfield  {title}
  {\bibinfo {title} {{Identification of ferroelectric domain structure
  sensitive phonon modes in potassium titanyl phosphate: A fundamental
  study}},\ }\href {https://doi.org/10.1063/1.4940964} {\bibfield  {journal}
  {\bibinfo  {journal} {Journal of Applied Physics}\ }\textbf {\bibinfo
  {volume} {119}},\ \bibinfo {pages} {044103} (\bibinfo {year}
  {2016})}\BibitemShut {NoStop}%
\bibitem [{\citenamefont {R{\"{u}}sing}\ \emph {et~al.}(2018)\citenamefont
  {R{\"{u}}sing}, \citenamefont {Neufeld}, \citenamefont {Brockmeier},
  \citenamefont {Eigner}, \citenamefont {Mackwitz}, \citenamefont {Spychala},
  \citenamefont {Silberhorn}, \citenamefont {Schmidt}, \citenamefont {Berth},
  \citenamefont {Zrenner},\ and\ \citenamefont {Sanna}}]{Rusing2018}%
  \BibitemOpen
  \bibfield  {author} {\bibinfo {author} {\bibfnamefont {M.}~\bibnamefont
  {R{\"{u}}sing}}, \bibinfo {author} {\bibfnamefont {S.}~\bibnamefont
  {Neufeld}}, \bibinfo {author} {\bibfnamefont {J.}~\bibnamefont {Brockmeier}},
  \bibinfo {author} {\bibfnamefont {C.}~\bibnamefont {Eigner}}, \bibinfo
  {author} {\bibfnamefont {P.}~\bibnamefont {Mackwitz}}, \bibinfo {author}
  {\bibfnamefont {K.}~\bibnamefont {Spychala}}, \bibinfo {author}
  {\bibfnamefont {C.}~\bibnamefont {Silberhorn}}, \bibinfo {author}
  {\bibfnamefont {W.~G.}\ \bibnamefont {Schmidt}}, \bibinfo {author}
  {\bibfnamefont {G.}~\bibnamefont {Berth}}, \bibinfo {author} {\bibfnamefont
  {A.}~\bibnamefont {Zrenner}},\ and\ \bibinfo {author} {\bibfnamefont
  {S.}~\bibnamefont {Sanna}},\ }\bibfield  {title} {\bibinfo {title} {{Imaging
  of ${180}^{\ensuremath{\circ}}$ ferroelectric domain walls in uniaxial
  ferroelectrics by confocal Raman spectroscopy: Unraveling the contrast
  mechanism}},\ }\href {https://doi.org/10.1103/PhysRevMaterials.2.103801}
  {\bibfield  {journal} {\bibinfo  {journal} {Phys. Rev. Mater.}\ }\textbf
  {\bibinfo {volume} {2}},\ \bibinfo {pages} {103801} (\bibinfo {year}
  {2018})}\BibitemShut {NoStop}%
\bibitem [{\citenamefont {Reitzig}\ \emph {et~al.}(2021)\citenamefont
  {Reitzig}, \citenamefont {R{\"{u}}sing}, \citenamefont {Zhao}, \citenamefont
  {Kirbus}, \citenamefont {Mookherjea},\ and\ \citenamefont
  {Eng}}]{Reitzig2021}%
  \BibitemOpen
  \bibfield  {author} {\bibinfo {author} {\bibfnamefont {S.}~\bibnamefont
  {Reitzig}}, \bibinfo {author} {\bibfnamefont {M.}~\bibnamefont
  {R{\"{u}}sing}}, \bibinfo {author} {\bibfnamefont {J.}~\bibnamefont {Zhao}},
  \bibinfo {author} {\bibfnamefont {B.}~\bibnamefont {Kirbus}}, \bibinfo
  {author} {\bibfnamefont {S.}~\bibnamefont {Mookherjea}},\ and\ \bibinfo
  {author} {\bibfnamefont {L.~M.}\ \bibnamefont {Eng}},\ }\bibfield  {title}
  {\bibinfo {title} {{“Seeing Is Believing”—In-Depth Analysis by
  Co-Imaging of Periodically-Poled X-Cut Lithium Niobate Thin Films}},\
  }\bibfield  {journal} {\bibinfo  {journal} {Crystals}\ }\textbf {\bibinfo
  {volume} {11}},\ \href {https://doi.org/10.3390/cryst11030288}
  {10.3390/cryst11030288} (\bibinfo {year} {2021})\BibitemShut {NoStop}%
\bibitem [{\citenamefont {Stone}\ and\ \citenamefont
  {Dierolf}(2012)}]{Stone2012}%
  \BibitemOpen
  \bibfield  {author} {\bibinfo {author} {\bibfnamefont {G.}~\bibnamefont
  {Stone}}\ and\ \bibinfo {author} {\bibfnamefont {V.}~\bibnamefont
  {Dierolf}},\ }\bibfield  {title} {\bibinfo {title} {{Influence of
  ferroelectric domain walls on the Raman scattering process in lithium
  tantalate and niobate}},\ }\href {https://doi.org/10.1364/OL.37.001032}
  {\bibfield  {journal} {\bibinfo  {journal} {Opt. Lett.}\ }\textbf {\bibinfo
  {volume} {37}},\ \bibinfo {pages} {1032} (\bibinfo {year}
  {2012})}\BibitemShut {NoStop}%
\end{thebibliography}

%

\end{document}